\documentclass[%
 reprint,
 amsmath,amssymb,
aps
pra,twocolumn,
]{revtex4-2}
\usepackage{braket}
\usepackage{graphicx}
\usepackage{bm}
\usepackage{float}
\usepackage{color}
\usepackage{soul}
\usepackage[caption = false]{subfig}
\usepackage{hyperref}
\usepackage{placeins}

\usepackage{dsfont}
\usepackage[normalem]{ulem}

\begin{document}

\title{Limitations of probabilistic error cancellation for open dynamics beyond sampling overhead}
\author{Yue Ma$^1$}
\author{M. S. Kim$^1$}

\affiliation{$^1$QOLS, Blackett Laboratory, Imperial College London, London SW7 2AZ, United Kingdom\\
}

\begin{abstract}

Quantum simulation of dynamics is an important goal in the NISQ era, within which quantum error mitigation may be a viable path towards modifying or eliminating the effects of noise. Most studies on quantum error mitigation have been focused on the resource cost due to its exponential scaling in the circuit depth. Methods such as probabilistic error cancellation rely on discretizing the evolution into finite time steps and applying the mitigation layer after each time step, modifying only the noise part without any Hamiltonian-dependence. This may lead to Trotter-like errors in the simulation results even if the error mitigation is implemented ideally, which means that the number of samples is taken as infinite. Here we analyze the aforementioned errors which have been largely neglected before. We show that, they are determined by the commutating relations between the superoperators of the unitary part, the device noise part and the noise part of the open dynamics to be simulated. We include both digital quantum simulation and analog quantum simulation setups, and consider defining the ideal error mitigation map both by exactly inverting the noise channel and by approximating it to the first order in the time step. We take single-qubit toy models to numerically demonstrate our findings. Our results illustrate fundamental limitations of applying probabilistic error cancellation in a stepwise manner to continuous dynamics, thus motivating the investigations of truly time-continuous error cancellation methods.

\end{abstract}

\date{\today}

\maketitle

\section{introduction}

Dynamics of quantum systems are unavoidably noisy due to the interaction with the environment. Elements of quantum computers, therefore, are also subject to a variety of noises. The main techniques of reducing the effects of noise belong to two categories, quantum error correction~\cite{terhal2015quantum} and quantum error mitigation~\cite{cai2022quantum}. Quantum error correction encodes the quantum information in a much larger Hilbert space in order to achieve fault-tolerance. Quantum error mitigation, on the contrary, makes use of a large number of copies of the quantum system and some knowledge of the noise structures in order to achieve a noise-free ensemble average result via statistical post-processing. Given the current and near-future error rates for a single quantum register~\cite{preskill2018quantum}, the implementations of various quantum error mitigation methods on actual quantum hardware~\cite{cai2022quantum} are considered to be an important milestone on the way to the long-term goal of fault-tolerant quantum computing.

One important application of quantum computing before full fault-tolerance can be achieved, is quantum simulation, especially the simulation of quantum dynamics~\cite{daley2022practical}. Indeed, different quantum error mitigation methods have been applied to different implementations of quantum computers to simulate unitary dynamics~\cite{kim2023scalable,van2023probabilistic,arute2020observation,rosenberg2023dynamics,kim2023evidence}, with the possible potential of showing quantum advantages over classical computers. Recently, it has been proposed that quantum error mitigation can be applied to partially mitigate the noise of the quantum computer, so that open dynamics can be simulated, which have applications in many aspects in quantum chemistry and quantum biology~\cite{guimaraes2023noise}.

Although widely used, methods of quantum error mitigation are known to suffer from an exponential scaling of the sampling overhead~\cite{takagi2022fundamental,takagi2022universal,tsubouchi2022universal}, and to what extent this can be optimized remains an open question. On the other hand, as quantum error mitigation aims at removing noise, how its effects change with different unitary parts of the dynamics is rarely explored, partially due to the exponential sampling cost mentioned above that poses significant numerical challenge. This is especially important for simulating the dynamics, as according to the theory of open dynamics~\cite{breuer2002theory,gardiner2004quantum}, in general it is impossible to find a Hamiltonian-independent map that precisely converts a noisy time evolution to a noiseless one.

In this paper, we focus on the simulation errors of open and closed dynamics when probabilistic error cancellation, which is one of the quantum error mitigation methods, is applied to reshape the noise distribution of a noisy quantum simulation process. In particular, we assume that the number of samples is infinitely large, such that the probabilistic error cancellation is implemented in an ideal way. We describe a general framework to show that the simulation errors depend on the commutators between different superoperators governing the dynamics, and the contributing commutators are different depending on whether the simulation setup is digital or analog. We then illustrate our results numerically by taking a single-qubit toy model as an example, where analytical expressions are possible and the large sampling overhead for quantum error mitigation is not a limiting factor. Although the simulation errors we investigate here are unlikely to dominate over the numerical errors introduced by the sampling overhead for realistic models, our work points out the intrinsic limitations of combining step-wise protocols to continuous dynamics, which can only be overcome by applying time-continuous protocols instead.

\section{the general framework}\label{sec:general}

We consider the simulation of a Lindblad master equation that describes weak couplings between the system and a Markovian environment~\cite{breuer2002theory,milz2017introduction},
\begin{align}
    \frac{d\rho(t)}{dt}&=-i[\hat{H},\rho(t)]\nonumber\\    &+\sum_k\gamma_k\Big(\hat{L}_k\rho(t)\hat{L}_k^{\dagger}-\frac{1}{2}\big(\hat{L}_k^{\dagger}\hat{L}_k\rho(t)+\rho(t)\hat{L}_k^{\dagger}\hat{L}_k\big)\Big),
\end{align}
where $\hat{H}$ is the Hamiltonian corresponding to the unitary dynamics of the system, $\hat{L}_k$ are Lindblad operators with error rate $\gamma_k$, For simplicity, we have taken $\hbar=1$, and have assumed that $\hat{H}$ and $\hat{L}_k$ are time-independent. We rewrite the master equation into the superoperator form by vectorizing the $d\times d$ density matrix $\rho$ into the column-ordered $d^2\times1$ vector $\boldsymbol{\rho}$~\cite{campaioli2023tutorial}
\begin{align}
    &\frac{d\boldsymbol{\rho}(t)}{dt}=(\mathcal{L}_h+\mathcal{L}_d)\boldsymbol{\rho}(t),\label{eq:meq}\\
    &\mathcal{L}_h=-i(\mathds{1}\otimes\hat{H}-\hat{H}^T\otimes\mathds{1}),\\    &\mathcal{L}_d=\sum_k\gamma_k\Big(\hat{L}_k^{*}\otimes\hat{L}_k-\frac{1}{2}\big(\mathds{1}\otimes\hat{L}_k^{\dagger}\hat{L}_k+\hat{L}_k^{T}\hat{L}_k^{*}\otimes\mathds{1}\big)\Big),
\end{align}
where $\mathcal{L}_h$ is the superoperator describing the unitary part of the dynamics governed by the Hamiltonian, and $\mathcal{L}_d$ is the superoperator describing the noise part of the open dynamics to be simulated. Both $\mathcal{L}_h$ and $\mathcal{L}_d$ are $d^2\times d^2$ matrices acting on the vectorized density matrix $\boldsymbol{\rho}$ from the left. As we have assumed that $\hat{H}$ and $\hat{L}_k$ are time-independent, $\mathcal{L}_h$ and $\mathcal{L}_d$ are time-independent as well. The solution of Eq.~\eqref{eq:meq} is therefore
\begin{align}\label{eq:vec_rho}
    \boldsymbol{\rho}(t)=e^{\mathcal{L}_ht+\mathcal{L}_dt}\boldsymbol{\rho}(0).
\end{align}
Note that, according to the Baker-Campbell-Hausdorff formula~\cite{rossmann2006lie}, $\exp(\mathcal{L}_ht+\mathcal{L}_dt)=\exp(\mathcal{L}_dt)\exp(\mathcal{L}_ht)$ works only if $\mathcal{L}_d$ and $\mathcal{L}_h$ commute. Also note that the case of simulating closed dynamics can be included by simply taking $\mathcal{L}_d=0$.

Typically, a quantum computer aims at simulating the unitary dynamics $\mathcal{L}_h$, but unavoidable sources of noise imply that the quantum computer is in fact simulating some sort of open dynamics. We consider applying quantum error mitigation to attenuate the noise of the quantum computer, such that effectively we can simulate the dynamics governed by $\mathcal{L}_h+\mathcal{L}_d$ rather than the intrinsic noise of the quantum computer. Specifically, we choose the method of probabilistic error cancellation~\cite{temme2017error}. It aims at implementing a map that cancels the effect of the noise channel. As this map is not physical, it can only be realised statistically. A large number of trials are required. Within each trial, some unitary operators are sampled and applied to the state. Taking the statistical average over all the trials, together with classical post-processing, will effectively implement the desired non-physical map. We will explicitly illustrate these steps in Sec.~\ref{sec:PEC} with the example of a single qubit. As probabilistic error cancellation is designed to act in discrete time steps~\cite{van2023probabilistic,guimaraes2023noise,sun2021mitigating}, we consider the time interval $\Delta t$, such that the state of the system at time $t+\Delta t$ is given by $\mathcal{M}\mathcal{C}\boldsymbol{\rho}(t)$, where $\mathcal{M}$ is the superoperator corresponding to the non-physical map to be realized by probabilistic error cancellation, and $\mathcal{C}$ is the superoperator describing the noisy implementation of the unitary dynamics during the time step $\Delta t$. The form of $\mathcal{C}$ depends on whether the quantum computer is digital or analog, and will be discussed in the next subsections. The error mitigation superoperator $\mathcal{M}$ should be designed such that 
\begin{align}\label{eq:M}
    \mathcal{M}\mathcal{C}=e^{\mathcal{L}_h\Delta t+\mathcal{L}_d\Delta t}.
\end{align}
That is, we want to simulate open dynamics on a noisy quantum computer, which can be executed by mitigating the noisy unitary evolution such that effectively the remaining noise matches the noise coming from the interactions with the environment that are to be simulated. As a special case, this also includes the simulation of closed quantum systems. We are going to show that, if we require $\mathcal{M}$ to be independent of $\mathcal{L}_h$, which is a standard assumption as error mitigation protocols only aim at cancelling the noise, unless special conditions are satisfied, in general Eq.~\eqref{eq:M} cannot hold precisely. The errors resemble Trotter errors~\cite{rossmann2006lie}, but are in terms of the superoperators corresponding to the unitary part and noisy part of the dynamics.

\subsection{digital quantum simulation}\label{sec:digitalGeneral}

For a digital quantum computer, the noisy implementation of a unitary gate is usually modelled as a unitary layer followed by a layer of the noise channel~\cite{van2023probabilistic,guimaraes2023noise}. Specifically,
\begin{align}\label{eq:dqsC}
   \mathcal{C}=e^{\mathcal{L}_n\Delta t}e^{\mathcal{L}_h\Delta t},
\end{align}
where $\mathcal{L}_n$ is the superoperator describing the device noise channel.

In some cases the device noise channel is defined by parameters that have already taken into account the exponential over $\Delta t$. This will be discussed in detail with the numerical examples in Sec.~\ref{sec:digLambda}. Assuming that $\exp(\mathcal{L}_n\Delta t)$ is invertible, the error mitigation superoperator $\mathcal{M}$ should satisfy
\begin{align}
    \mathcal{M}=e^{\mathcal{L}_d\Delta t+\mathcal{L}_h\Delta t}e^{-\mathcal{L}_h\Delta t}e^{-\mathcal{L}_n\Delta t}.
\end{align}
For the simulation of closed dynamics, $\mathcal{L}_d=0$. We therefore require
\begin{align}\label{eq:Mdig}
    \mathcal{M}=e^{-\mathcal{L}_n\Delta t}.
\end{align}
This is independent of the unitary part $\mathcal{L}_h$, and only cancels the device noise. On the other hand, for the simulation of open dynamics~\cite{guimaraes2023noise}, $\mathcal{L}_d\neq0$. Different forms of $\mathcal{M}$ are resulted in, depending on the commutator between the superoperators $\mathcal{L}_d$ and $\mathcal{L}_h$. If $[\mathcal{L}_d,\mathcal{L}_h]=0$, we have
\begin{align}\label{eq:Mdn}
    \mathcal{M}=e^{(\mathcal{L}_d-\mathcal{L}_n)\Delta t}.
\end{align}
The error mitigation superoperator $\mathcal{M}$ is still independent of $\mathcal{L}_h$. Note that, satisfying the relation $[\mathcal{L}_d,\mathcal{L}_h]=0$ may pose a requirement on the form of $\mathcal{L}_h$, but for special $\mathcal{L}_d$ such as depolarizing noise~\cite{van2023probabilistic}, $[\mathcal{L}_d,\mathcal{L}_h]=0$ holds for any Hamiltonian. Finally, if $[\mathcal{L}_d,\mathcal{L}_h]\neq0$, according to the Zassenhaus formula~\cite{campaioli2023tutorial},
\begin{align}\label{eq:Mcmt}
    \mathcal{M}=e^{-\frac{1}{2}(\Delta t)^2[\mathcal{L}_d,\mathcal{L}_h]}e^{(\mathcal{L}_d-\mathcal{L}_n)\Delta t}\cdots,
\end{align}
where we have neglected terms that contain high orders of $\Delta t$ in the exponential. The first term on the right hand side of Eq.~\eqref{eq:Mcmt} implies that $\mathcal{M}$ has to depend on the Hamiltonian $\hat{H}$. Ignoring this term and the higher order terms will lead to Trotter errors, which depend on the superoperator commutator $[\mathcal{L}_d,\mathcal{L}_h]$ and can be reduced by decreasing the time step $\Delta t$. However, the large sampling overhead of probabilistic error cancellation may put a lower bound on the implementable values of $\Delta t$.

\subsection{analog quantum simulation}\label{sec:analog}

For an analog quantum computer, it is no longer possible to separate the unitary dynamics from the device noise. Instead. both of them act on the system simultaneously~\cite{sun2021mitigating}, namely, we now have
\begin{align}
   \mathcal{C}=e^{\mathcal{L}_n\Delta t+\mathcal{L}_h\Delta t}.
\end{align}
The error mitigation superoperator $\mathcal{M}$ is thus required to be
\begin{align}
    \mathcal{M}=e^{\mathcal{L}_d\Delta t+\mathcal{L}_h\Delta t}e^{-\mathcal{L}_n\Delta t-\mathcal{L}_h\Delta t}.
\end{align}
Using Baker-Campbell-Hausdorff formula, we can rewrite $\mathcal{M}$ as
\begin{align}
    \mathcal{M}=e^{-\frac{1}{2}(\Delta t)^2[\mathcal{L}_d-\mathcal{L}_n,\mathcal{L}_h]}e^{(\mathcal{L}_d-\mathcal{L}_n)\Delta t}\cdots,
\end{align}
where $\cdots$ represents the terms coming from commutating relations that have higher orders of $\Delta t$ in the exponential. Note that in this case, only if $\mathcal{L}_d-\mathcal{L}_n$ commutes with $\mathcal{L}_h$ will $\mathcal{M}$ be Hamiltonian-independent without Trotter error, i.e., $\mathcal{M}$ is given by Eq.~\eqref{eq:Mdn}. This is different from the case of a digital quantum computer where the device noise $\mathcal{L}_n$ does not contribute to the commutating relation. Thus one important consequence for an analog quantum computer is that, \textit{for simulations of closed dynamics $\mathcal{L}_d=0$, if $[\mathcal{L}_n,\mathcal{L}_h]\neq0$, it is impossible to precisely recover the unitary dynamics in a Hamiltonian-independent way}. This no-go theorem applies to any protocol aiming at cancelling the noise in a dynamical process, including different methods of quantum error mitigation, such as the method described in Ref.~\cite{sun2021mitigating}.

\section{toy models}\label{sec:toyModel}

In this section we take single-qubit toy models to demonstrate the various simulation errors we have discussed in Sec.~\ref{sec:general}. Although oversimplified, single-qubit models have certain advantages, a few of which are the following. The unitary dynamics can be expressed exactly as $\exp(-i\hat{H}t)$ without Trotter errors, so that we can focus on the Trotter errors related to the noise superoperators. Due to the small Hilbert space dimension, analytical solutions can be kept track of. The number of samples for realizing probabilistic error cancellation can be very large as the numerical cost per sample is very low.

We consider the Hamiltonian to be
\begin{align}\label{eq:Hamiltonian}
    \hat{H}=\omega(\sin\beta\hat{X}-\cos\beta\hat{Y}),
\end{align}
where the Pauli operators are $\hat{X}=|0\rangle\langle1|+|1\rangle\langle0|$, $\hat{Y}=i|0\rangle\langle1|-i|1\rangle\langle0|$, $\hat{Z}=|1\rangle\langle1|-|0\rangle\langle0|$, and we have assumed $|1\rangle=(1,0)^T$ and $|0\rangle=(0,1)^T$. The initial state is chosen to be $\rho(0)=|1\rangle\langle1|$. Physically, this describes the Rabi oscillation of a two-level quantum system initialised to its excited state, as will be described in detail later. The parameter $\beta$ can be changed to obtain different commutators between the superoperators. In the following subsections, we will consider the simulations of closed dynamics and open dynamics separately. Within each category, we discuss both the digital simulation setup and the analog simulation setup, including different device noise models. For simplicity, we focus on Pauli noise models for both the device noise and the noise part of the open dynamics, so that the error mitigation superoperator Eq.~\eqref{eq:Mdn} has a closed form expression~\cite{van2023probabilistic}. 
Specifically, the Lindblad operators $\hat{L}_k$ are proportional to the single qubit Pauli operators (see Appendix~\ref{sec:PauliDetails}).
Moreover, we consider both the exact implementation of Eq.~\eqref{eq:Mdn} and approximate implementations where Eq.~\eqref{eq:Mdn} is expanded to the first order in $\Delta t$. Due to the Trotter errors that we have discussed, applying $\mathcal{M}$ according to Eq.~\eqref{eq:Mdn} for each time step may not lead to the accurate simulation result. We therefore use the fidelity to quantify the simulation errors.

\subsection{probabilistic error cancellation}\label{sec:PEC}

The quantum error mitigation method of probabilistic error cancellation~\cite{temme2017error} uses multiple copies of the quantum system to stochastically implement a non-physical map. For a single qubit as described above, the error mitigation superoperator Eq.~\eqref{eq:Mdn} is in the following form,
\begin{equation}\label{eq:MsingleQ}
    \mathcal{M}=q_0\mathds{1}\otimes\mathds{1}+q_1\hat{X}\otimes\hat{X}+q_2\hat{Y}^{*}\otimes\hat{Y}+q_3\hat{Z}\otimes\hat{Z},
\end{equation}
where $q_0+q_1+q_2+q_3=1$ and $q_0>0$ (see Appendix~\ref{sec:PauliDetails} for details). As $\mathcal{M}$ might convert a time evolution to a less noisy version, it might be non-physical, corresponding to negative values of $q_1$, $q_2$ and $q_3$. In this case, Eq.~\eqref{eq:MsingleQ} can only be implemented stochastically, in combination with post processing. To be specific, we first define sampling probabilities,
\begin{align}
    \mu_1=\frac{|q_1|}{q_0+|q_1|+|q_2|+|q_3|},\label{eq:mu1}\\
    \mu_2=\frac{|q_2|}{q_0+|q_1|+|q_2|+|q_3|},\label{eq:mu2}\\
    \mu_3=\frac{|q_3|}{q_0+|q_1|+|q_2|+|q_3|},\label{eq:mu3}
\end{align}
where $|\cdot|$ refers to taking the absolute value.
For each sample, a unitary operator is applied to the qubit, and a classical prefactor is multiplied to the density matrix. With probability $1-\mu_1-\mu_2-\mu_3$, the unitary operator is the identity operator $\mathds{1}$ and the prefactor is $q_0+|q_1|+|q_2|+|q_3|$. With probability $\mu_i$ ($i=1,2,3$), the unitary operator is the Pauli operator $\hat{P}_i$ ($\hat{P}_1=\hat{X}$, $\hat{P}_2=\hat{Y}$, $\hat{P}_3=\hat{Z}$) and the prefactor is $\mathrm{sign}(q_i)\times(q_0+|q_1|+|q_2|+|q_3|)$, where $\mathrm{sign(\cdot)}$ refers to taking the sign. Averaging over an infinite number of samples corresponds to the ideal implementation of Eq.~\eqref{eq:MsingleQ}. For practical numerical simulations, however, there are finite errors that decrease with an increasing sample size. In Appendix~\ref{sec:sampling}, we use numerical examples to show that deviations from the ideal sampling probabilities Eqs.~\eqref{eq:mu1}-\eqref{eq:mu3} may lead to non-physical simulation results, as the post-processing is not a physical process.

\begin{figure}[t]
\centering
\includegraphics[width=0.48\textwidth]{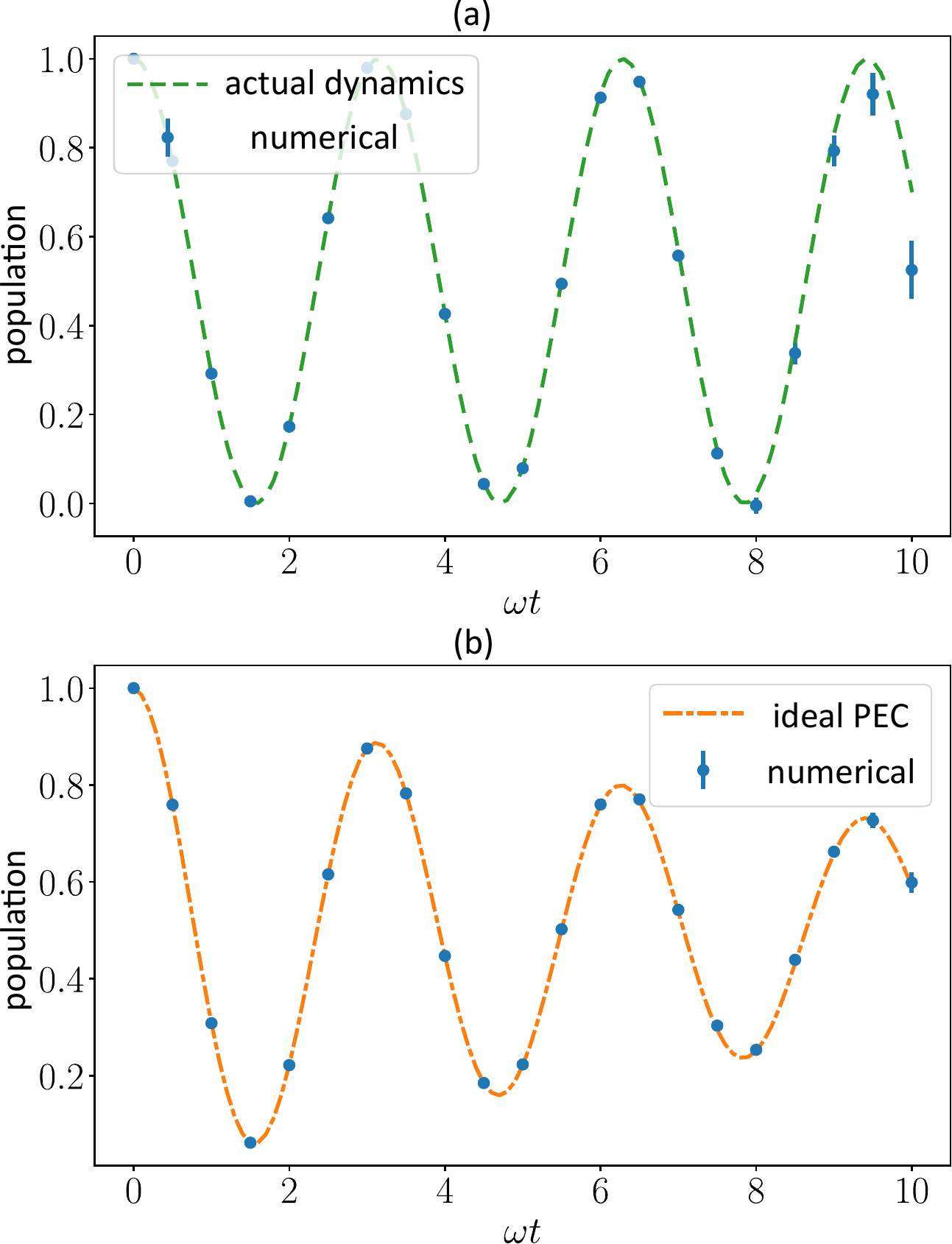}
\caption{Combining step-wise probabilistic error cancellation with a noisy digital quantum computer to simulate closed dynamics. We have chosen the parameters as $\omega=1$, $\Delta t=0.5$, $\lambda_1=\lambda_2=\lambda_3=0.05$. The blue numerical data points are the ensemble averages over $2\times10^7$ samples, and the errorbars correspond to the standard deviation. The green dashed line is the actual closed dynamics to be simulated, following Eq.~\eqref{eq:Csol}. The orange dot-dashed line is the analytical result for the ideal implementation of the error mitigation maps, i.e., in the limit of an infinitely large sample size. (a) Exact probabilistic error cancellation. The parameters $q_0$, $q_1$, $q_2$, $q_3$ in Eq.~\eqref{eq:MsingleQ} follow the exact inversion of the noise channel, as given in Eqs.~\eqref{eq:qED1}-\eqref{eq:qED3}. Here the green dashed line also corresponds to the ideal implementation of probabilistic error cancellation, therefore there is no systematic error. (b) Approximate probabilistic error cancellation. The parameters $q_0$, $q_1$, $q_2$, $q_3$ are the approximate ones following Eq.~\eqref{eq:MdigitalApprox}. The orange dot-dashed line follows Eq.~\eqref{eq:rho11}. It implies that even in the limit of an infinitely large number of samples, the noise cannot be fully cancelled.}\label{fig:DCEA}
\end{figure}

\subsection{closed dynamics}

The desired dynamics to be simulated are determined by the Hamiltonian Eq.~\eqref{eq:Hamiltonian}. The time evolution of the population of the excited state can be expressed analytically independent of the parameter $\beta$,
\begin{align}\label{eq:Csol}
    \langle 1|\rho(t)|1\rangle=\frac{1}{2}\Big(1+\cos(2\omega t)\Big).
\end{align}
It represents an oscillation between $0$ and $1$ with period $\pi/\omega$, as shown in the green dashed line in Fig.~\ref{fig:DCEA}(a).

\subsubsection{digital quantum simulation}\label{sec:digLambda}

We first consider a noisy digital quantum simulation of the closed dynamics. As discussed in Sec.~\ref{sec:digitalGeneral}, the noise is typically modelled by a separate layer of noise channel following the unitary layer within each time step. We assume that the noise channel is expressed as
\begin{equation}\label{eq:digitalNoise}
    \mathcal{N}(\rho)=(1-\lambda_1-\lambda_2-\lambda_3)\rho+\lambda_1\hat{X}\rho\hat{X}+\lambda_2\hat{Y}\rho\hat{Y}+\lambda_3\hat{Z}\rho\hat{Z},
\end{equation}
where the coefficients satisfy $\lambda_1,\lambda_2,\lambda_3\geq0$, $\lambda_1+\lambda_2+\lambda_3\leq1$, so that $\mathcal{N}(\rho)$ is a physical map. The error mitigation layer aims at cancelling the effect of $\mathcal{N}(\rho)$. Rigorous calculation steps involve finding $\mathcal{L}_n$ expressed via $\lambda_1,\lambda_2,\lambda_3$, and then calculating the error mitigation superoperator following Eq.~\eqref{eq:Mdig}. This will be described in details in Sec.~\ref{sec:odq} when we consider the digital simulation of open dynamics. For the case here, we can directly take the inverse matrix of the superoperator form of $\mathcal{N}$, and get $\mathcal{M}$ expressed in the form of Eq.~\eqref{eq:MsingleQ} with coefficients
\begin{align}
    q_1=&\frac{1}{4}\Big(1-\big(-\frac{1}{1-2\lambda_2-2\lambda_3}+\frac{1}{1-2\lambda_1-2\lambda_3}\label{eq:qED1}\nonumber\\
    &+\frac{1}{1-2\lambda_1-2\lambda_2}\big)\Big),\\
    q_2=&\frac{1}{4}\Big(1-\big(\frac{1}{1-2\lambda_2-2\lambda_3}-\frac{1}{1-2\lambda_1-2\lambda_3}\label{eq:qED2}\nonumber\\
    &+\frac{1}{1-2\lambda_1-2\lambda_2}\big)\Big),\\
    q_3=&\frac{1}{4}\Big(1-\big(\frac{1}{1-2\lambda_2-2\lambda_3}+\frac{1}{1-2\lambda_1-2\lambda_3}\label{eq:qED3}\nonumber\\
    &-\frac{1}{1-2\lambda_1-2\lambda_2}\big)\Big).
\end{align}
These define the exact probabilistic error cancellation map. Note that, however, the calculation requires inverting a large matrix, thus not scalable to more qubits. Moreover, more advanced methods are required to generalize to noise models that are not Pauli~\cite{temme2017error,endo2018practical}. We may therefore also consider an approximate implementation of Eq.~\eqref{eq:Mdig} if $\lambda_1,\lambda_2,\lambda_3\ll 1$,
\begin{equation}\label{eq:MdigitalApprox}
    q_1'=-\lambda_1,\ q_2'=-\lambda_2,\ q_3'=-\lambda_3,
\end{equation}
corresponding to negating the coefficients~\cite{guimaraes2023noise} in Eq.~\eqref{eq:digitalNoise}. It also corresponds to expanding Eqs.~\eqref{eq:qED1}-\eqref{eq:qED3} to the first order in $\lambda_1,\lambda_2,\lambda_3$. Coefficients in Eq.~\eqref{eq:MdigitalApprox} thus define the approximate probabilistic error cancellation map, which does not involve intensive calculations of matrix inversion.

As a numerical example, we consider $\beta=0$ in the Hamiltonian Eq.~\eqref{eq:Hamiltonian}, and take $\omega=1$. We choose the depolarizing noise channel, $\lambda_1=\lambda_2=\lambda_3=0.05$ in Eq.~\eqref{eq:digitalNoise}. We take the time step $\Delta t=0.5$ and $20$ layers of error mitigation. The results are shown in Fig.~\ref{fig:DCEA}. The numerical data come from sampling according to the probabilistic error cancellation procedure described in Sec.~\ref{sec:PEC}. The lines corresponding to the ideal implementations of probabilistic error cancellation, on the other hand, come from directly applying the superoperator Eq.~\eqref{eq:MsingleQ} to the vectorized density matrix $\boldsymbol{\rho}$, therefore equivalent to an infinite sample size. As discussed in Sec.~\ref{sec:digitalGeneral}, for the digital simulation of closed dynamics, Eq.~\eqref{eq:Mdig} fully recovers the unitary dynamics without any Trotter error, as Trotter errors come from the commutator $[\mathcal{L}_d,\mathcal{L}_h]$ and for closed dynamics $\mathcal{L}_d=0$. Thus in Fig.~\ref{fig:DCEA}(a), the ideal implementation of the exact probabilistic error cancellation coincides with the actual closed dynamics, both of which are described by Eq.~\eqref{eq:Csol}. The approximate probabilistic error cancellation plotted in Fig.~\ref{fig:DCEA}(b), on the other hand, is based on a linear approximation to Eq.~\eqref{eq:Mdig}. This approximation induces error in the simulation, which persists even in the limit of an infinite number of samples. In fact, the orange dot-dashed line in Fig.~\ref{fig:DCEA}(b) has an analytical closed-form expression due to the simplicity of the depolarizing noise channel we have chosen,
\begin{equation}\label{eq:rho11}
    \langle1|\rho(t)|1\rangle=\frac{1}{2}\Big(1+\big(1-16\lambda_{1}^2\big)^{t/\Delta t}\cos(2\omega t)\Big),
\end{equation}
where the time $t=N\Delta t$ is only defined at an integer multiple of $\Delta t$, $N=0,1,2,\cdots$. It is interesting to look at the limit of taking an infinitesimal time step, $\Delta t\rightarrow 0$. We can consider two distinct situations. For the first situation, we assume that $\lambda_1$ is independent of $\Delta t$. Taking the limit brings the right hand side of Eq.~\eqref{eq:rho11} to $1/2$. Features in the dynamics are completely lost as we have introduced a fixed amount of noise per step and we have an infinite number of steps. For the second situation, we assume that $\lambda_1$ is proportional to $\Delta t$. Taking the limit brings the right hand side of Eq.~\eqref{eq:rho11} to Eq.~\eqref{eq:Csol}, which is the closed dynamics. Details are included in Appendix~\ref{sec:digital}.

\subsubsection{analog quantum simulation}

Now we consider a noisy analog quantum simulation of the closed dynamics. In this case, the device noise acts simultaneously with the unitary dynamics to be simulated. As discussed in Sec.~\ref{sec:analog}, the Hamiltonian-independent error mitigation superoperator $\mathcal{M}$ is defined according to Eq.~\eqref{eq:Mdig}. However, there is no Trotter error only if $[\mathcal{L}_h,\mathcal{L}_n]=0$. We are going to consider two noise models. The first noise model is a depolarizing noise model, whose superoperator commutes with the superoperator of the Hamiltonian.  The second noise model only contains Pauli $\hat{X}$ operator, therefore only commutes with $\mathcal{L}_h$ if the parameter $\beta$ in the Hamiltonian Eq.~\eqref{eq:Hamiltonian} is taken as $\beta=\pi/2$. 

\begin{figure}[t]
\centering
\includegraphics[width=0.48\textwidth]{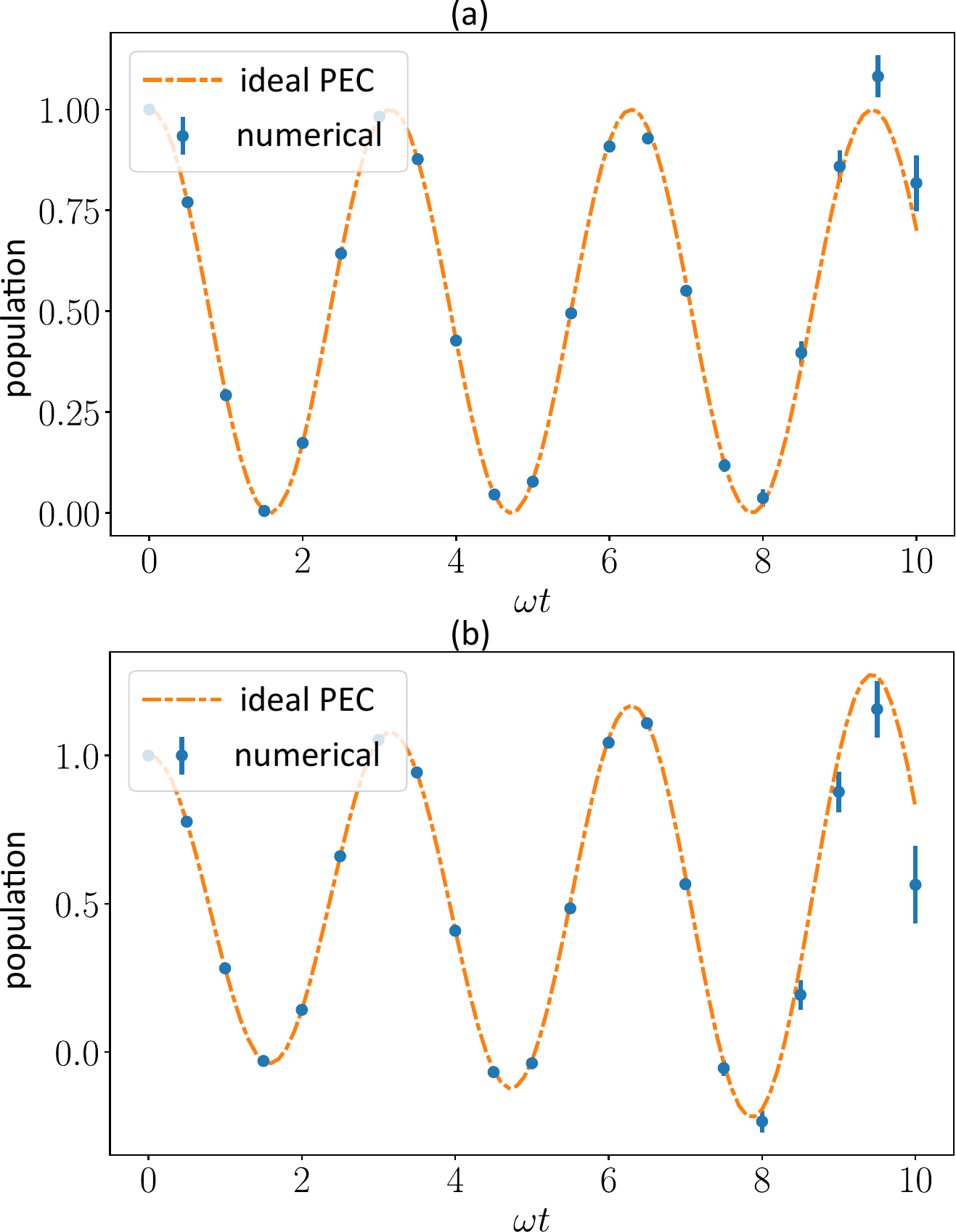}
\caption{Combining step-wise probabilistic error cancellation with a noisy analog quantum computer to simulate closed dynamics, assuming a depolarizing noise channel. We have chosen the parameters as $\omega=1$, $\Delta t=0.5$, $\kappa=0.1$. The blue numerical data points are the ensemble averages over $5\times10^6$ samples, and the errorbars correspond to the standard deviation. The orange dot-dashed line is the analytical result for the ideal implementation of the error mitigation maps, i.e., in the limit of an infinitely large sample size. (a) Exact probabilistic error cancellation. The parameters $q_0$, $q_1$, $q_2$, $q_3$ in Eq.~\eqref{eq:MsingleQ} correspond to the exact probabilistic error cancellation map, as given in Eq.~\eqref{eq:qACe}. Here the orange dot-dashed line also corresponds to the actual closed dynamics to be simulated, as $[\mathcal{L}_n,\mathcal{L}_h]=0$ implies no Trotter error. (b) Approximate probabilistic error cancellation. The parameters $q_0$, $q_1$, $q_2$, $q_3$ are the approximate ones following Eq.~\eqref{eq:qACa}. The orange dot-dashed line follows Eq.~\eqref{eq:rho_ee4}. It implies that even in the limit of an infinitely large number of samples, the result is different from the actual closed dynamics.}\label{fig:ACEA}
\end{figure}

\begin{figure}[t]
\centering
\includegraphics[width=0.48\textwidth]{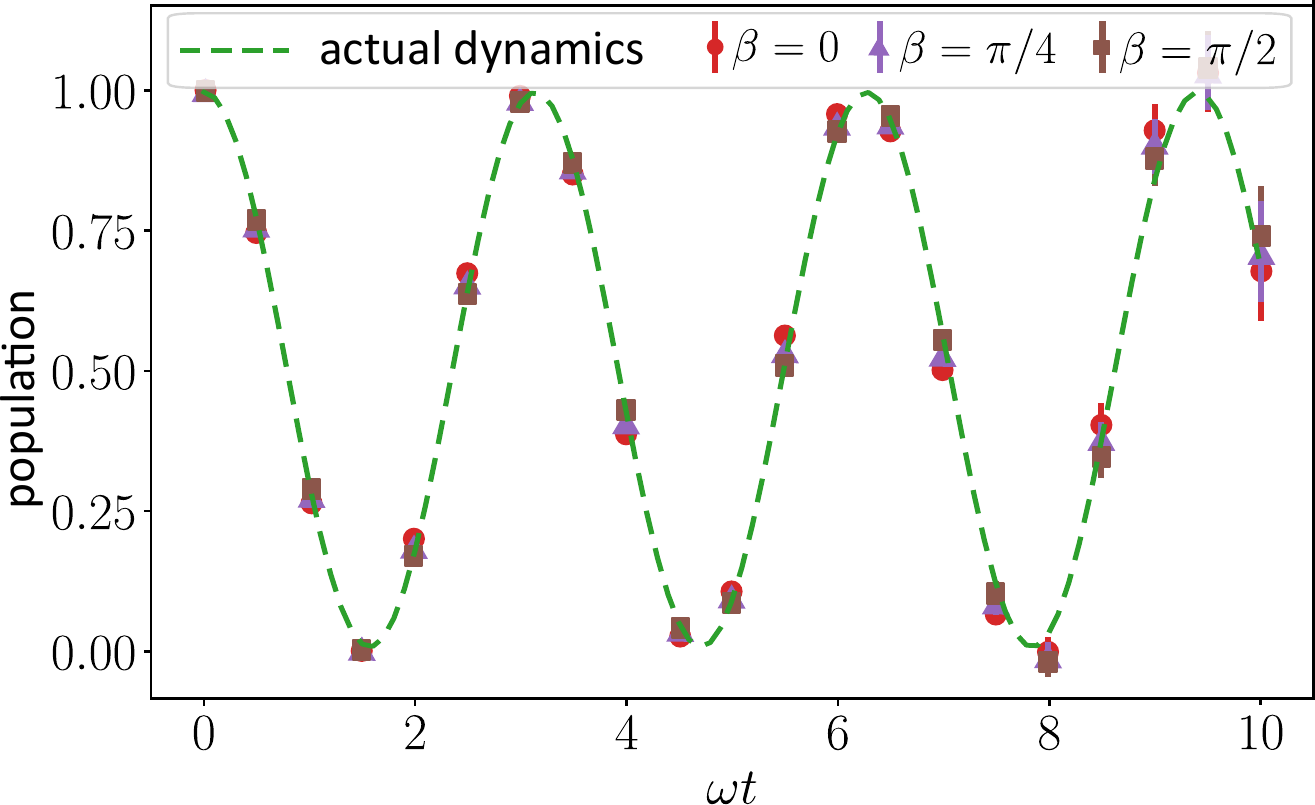}
\caption{Combining step-wise probabilistic error cancellation with a noisy analog quantum computer to simulate closed dynamics, assuming a Pauli-X noise channel. We have chosen the parameters as $\omega=1$, $\Delta t=0.5$, $\kappa=0.3$. The numerical data points are the ensemble averages over $5\times10^6$ samples, and the errorbars correspond to the standard deviations. The green dashed line is the actual closed dynamics to be simulated, as in Eq.~\eqref{eq:Csol}. The parameters $q_0$, $q_1$, $q_2$, $q_3$ in Eq.~\eqref{eq:MsingleQ} correspond to the exact probabilistic error cancellation map, as given in Eq.~\eqref{eq:qEX}. The brown square data points for $\beta=\pi/2$ correspond to the case without Trotter error. Decreasing $\beta$ to $\pi/4$ (purple triangle) and $0$ (red circle) leads to larger Trotter errors resulting from the non-zero $[\mathcal{L}_n,\mathcal{L}_h]$.}\label{fig:ACEX}
\end{figure}

\begin{figure}[t]
\centering
\includegraphics[width=0.48\textwidth]{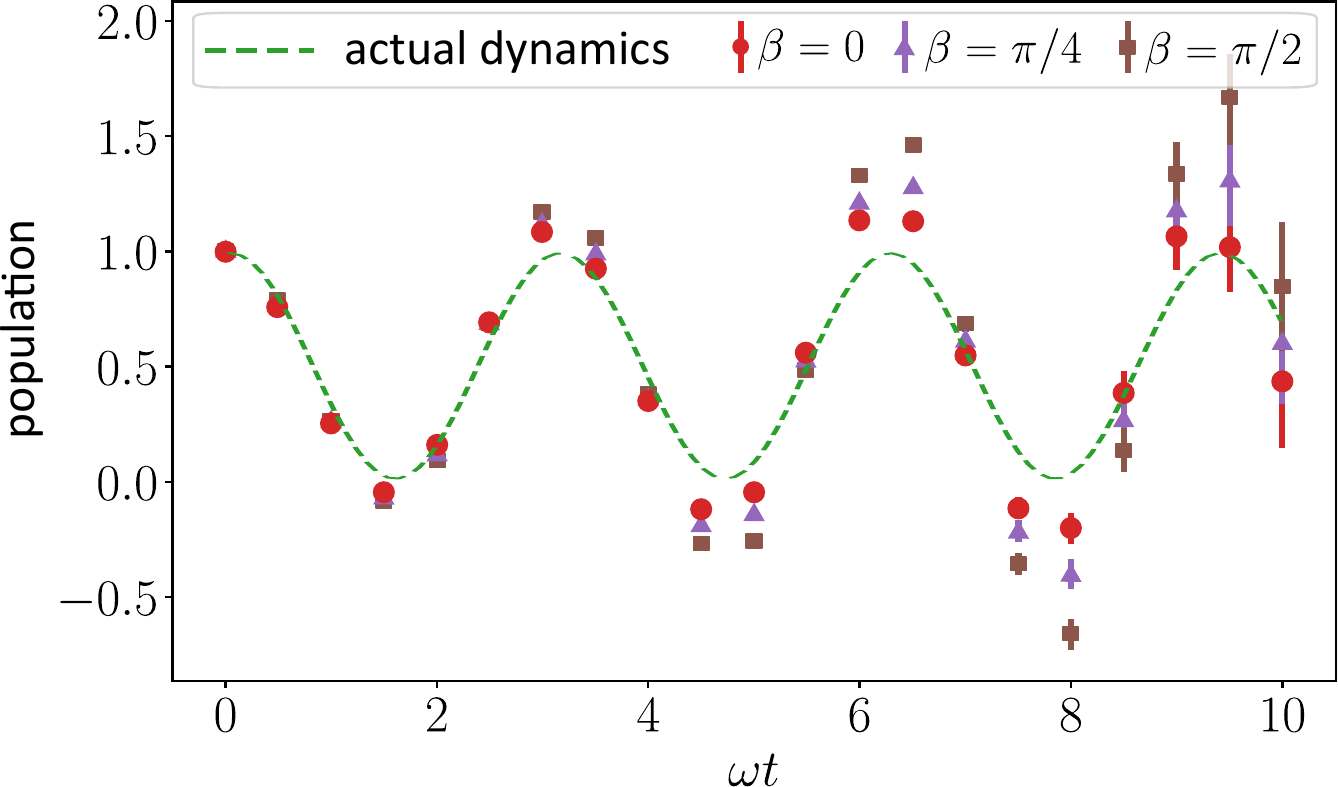}
\caption{Combining step-wise probabilistic error cancellation with a noisy analog quantum computer to simulate closed dynamics, assuming a Pauli-X noise channel. We have chosen the parameters as $\omega=1$, $\Delta t=0.5$, $\kappa=0.3$. The numerical data points are the ensemble averages over $5\times10^6$ samples, and the errorbars correspond to the standard deviations. The green dashed line is the actual closed dynamics to be simulated, as in Eq.~\eqref{eq:Csol}. The parameters $q_0$, $q_1$, $q_2$, $q_3$ in Eq.~\eqref{eq:MsingleQ} correspond to the approximate probabilistic error cancellation map, as given in Eq.~\eqref{eq:qAX}. Increasing $\beta$ from $0$ (red circle) to $\pi/4$ (purple triangle) to $\pi/2$ (brown square) leads to increasing simulation errors. Instead of Trotter errors, these errors are from the approximation of linear expansion, and can lead to non-physical results.}\label{fig:ACAX}
\end{figure}

\noindent\textbf{Depolarizing noise.} Under a depolarizing noise model, the quantum state evolves according to the master equation
\begin{align}\label{eq:master}
    \frac{d\rho(t)}{dt}=&-i[\hat{H},\rho(t)]+\kappa \Big(\hat{X}\rho(t)\hat{X}-\rho(t)\Big)\nonumber\\
    &+\kappa \Big(\hat{Y}\rho(t)\hat{Y}-\rho(t)\Big)+\kappa \Big(\hat{Z}\rho(t)\hat{Z}-\rho(t)\Big),
\end{align}
where the Hamiltonian $\hat{H}$ is given in Eq.~\eqref{eq:Hamiltonian}. The solution can be found to be
\begin{equation}\label{eq:rho_ee}
    \langle 1|\rho(t)|1\rangle=\frac{1}{2}\Big(1+e^{-4\kappa t}\cos(2\omega t)\Big),
\end{equation}
representing a Rabi oscillation with damping.
In order to apply probabilistic error cancellation, we discretize the continuous dynamics into time step $\Delta t$, and after each time step, apply the error mitigation step following Eq.~\eqref{eq:Mdig} and the procedure described in Sec.~\ref{sec:PEC}. Note that the device noise superoperator,
\begin{align}\label{eq:lnDp}
    \mathcal{L}_n&=\kappa(\hat{X}\otimes\hat{X}-\mathds{1}\otimes\mathds{1})+\kappa(\hat{Y}^*\otimes\hat{Y}-\mathds{1}\otimes\mathds{1})\nonumber\\
    &+\kappa(\hat{Z}\otimes\hat{Z}-\mathds{1}\otimes\mathds{1}), 
\end{align}
commutes with $\mathcal{L}_h$, therefore there is no Trotter error. As there are only Pauli errors, $\exp(\pm \mathcal{L}_n\Delta t)$ have simple closed form expressions~\cite{van2023probabilistic}. The error mitigation operator $\mathcal{M}$ is therefore expressed in the form of Eq.~\eqref{eq:MsingleQ} with coefficients
\begin{equation}\label{eq:qACe}
    q_1=q_2=q_3=\frac{1}{4}(1-e^{4\kappa\Delta t}).
\end{equation}
In addition to this exact map, we can also define an approximate error mitigation map by keeping only the linear term in $\Delta t$, although taking the approximation at different places lead to different expressions. Here we expand
$\exp(\mathcal{L}_n\Delta t)$
to the first order in $\Delta t$, and define the approximate $\mathcal{M}$ as the inverse of the expansion of $\exp(\mathcal{L}_n\Delta t)$, resulting in
\begin{equation}\label{eq:qACa}
    q_1'=q_2'=q_3'=-\frac{\kappa\Delta t}{1-4\kappa\Delta t}.
\end{equation}
We are going to introduce another way of linear expansion when we deal with the simulation of open dynamics in Sec.~\ref{sec:analogSub}.

In Fig.~\ref{fig:ACEA}(a) and (b), we show the numerical simulations following the coefficients in Eq.~\eqref{eq:qACe} and Eq.~\eqref{eq:qACa}, respectively. Note that the results are independent of the parameter $\beta$ in the Hamiltonian. While the exact coefficients Eq.~\eqref{eq:qACe} result in recovering the actual closed dynamics to be simulated in the limit of an infinite number of samples, the approximate coefficients Eq.~\eqref{eq:qACa} lead to errors that increase in time.
Note that these errors are not the Trotter error mentioned before. In fact, the dot-dashed orange line in Fig.~\ref{fig:ACEA}(b) has an analytical form,
\begin{equation}\label{eq:rho_ee4}
    \langle 1|\rho(t)|1\rangle=\frac{1}{2}\Big(1+\frac{e^{-4\kappa t}}{(1-4\kappa\Delta t)^{t/\Delta t}}\cos(2\omega t)\Big),
\end{equation}
defined for $t=N\Delta t$ with $N=0,1,2,\cdots$.
This corresponds to the solution of a master equation, Eq.~\eqref{eq:master} replacing $\kappa$ with
\begin{align}
    \kappa&\rightarrow\kappa+\frac{1}{4\Delta t}\log(1-4\kappa\Delta t)\nonumber\\
    &\approx\kappa\Big(-2\kappa\Delta t-\frac{16}{3}(\kappa\Delta t)^2+\cdots\Big).
\end{align}
Note that this is smaller than $0$, indicating a non-physical result. This is because the process of probabilistic error cancellation as described in Sec.~\ref{sec:PEC} involves post processing that does not satisfy physical rules.

\begin{figure}[t]
\centering
\includegraphics[width=0.48\textwidth]{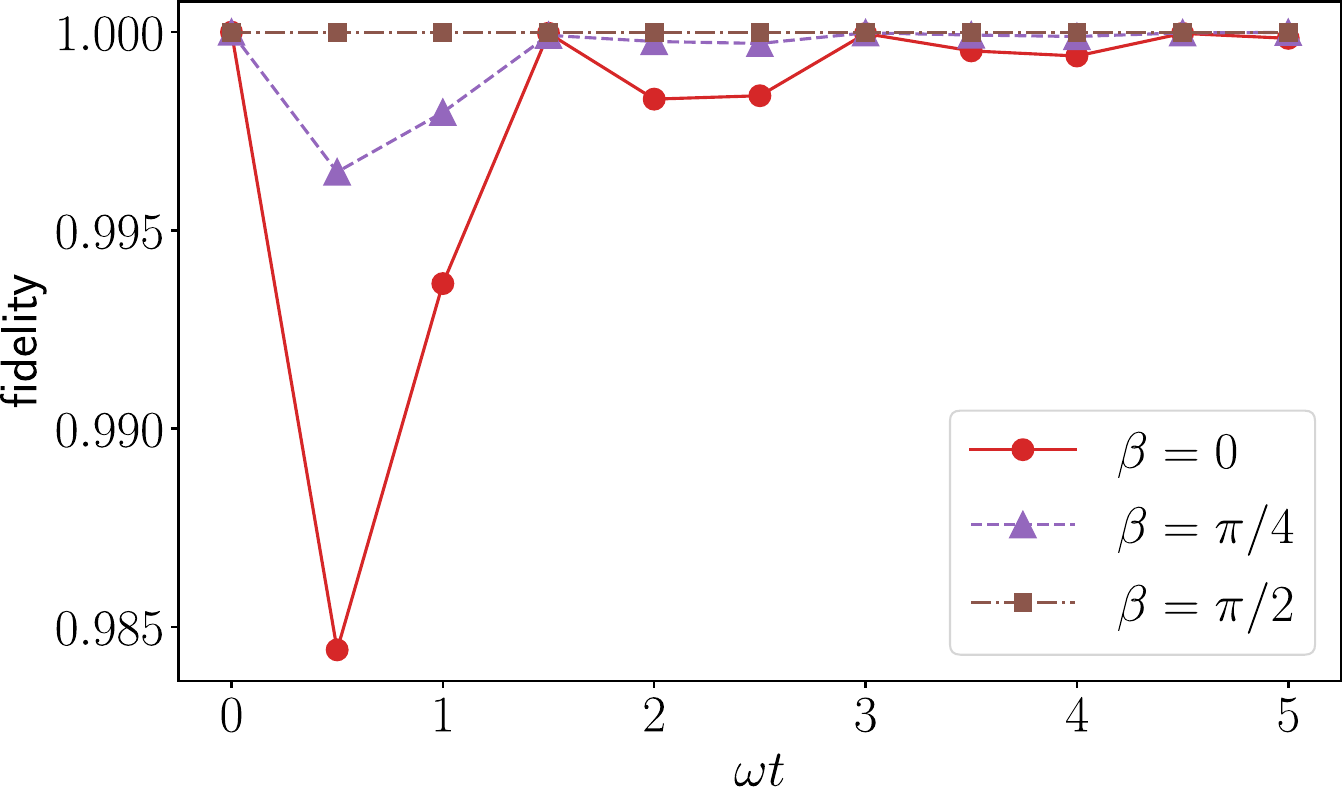}
\caption{Combining step-wise probabilistic error cancellation with a noisy digital quantum computer to simulate open dynamics. We have chosen the parameters as $\omega=1$, $\Delta t=0.5$, $\lambda_1=0.16$, $\lambda_2=0.12$, $\lambda_3=0.2$, $\gamma=0.3$. The fidelity is defined by comparing the exact solution of Eq.~\eqref{eq:GammamasterX} with the result from the ideal implementation (i.e., infinite sample size) of the exact error mitigation map with coefficients Eq.~\eqref{eq:qOED1}. The red solid line with circle marks is for $\beta=0$. The purple dashed line with triangle marks is for $\beta=\pi/4$. The brown dot-dashed line with square marks is for $\beta=\pi/2$, in which case there is no Trotter error, so that the fidelity remains $1$.
}\label{fig:DOE}
\end{figure}

\noindent\textbf{Pauli-X noise.} Next we consider a noise model that does not commute with the Hamiltonian Eq.~\eqref{eq:Hamiltonian} for general $\beta$. The evolution of the system follows
\begin{equation}\label{eq:masterX}
    \frac{d\rho(t)}{dt}=-i[\hat{H},\rho(t)]+\kappa \Big(\hat{X}\rho(t)\hat{X}-\rho(t)\Big)\nonumber.
\end{equation}
In this case, $\mathcal{L}_n=\kappa(\hat{X}\otimes\hat{X}-\mathds{1}\otimes\mathds{1})$ does not commute with $\mathcal{L}_h$ unless $\beta=\pi/2$. This also implies that a succinct form of $\langle 1|\rho(t)|1\rangle$ no longer exists.

For the exact probabilistic error cancellation map, we still define $\mathcal{M}$ according to Eq.~\eqref{eq:Mdig}. This corresponds to the coefficients in Eq.~\eqref{eq:MsingleQ} as
\begin{equation}\label{eq:qEX}
    q_1=\frac{1}{2}(1-e^{2\kappa\Delta t}),\ q_2=q_3=0.
\end{equation}
Numerical simulations are shown in Fig.~\ref{fig:ACEX}. Although not large, the deviation of the error mitigated results compared with the actual closed dynamics increases as $\beta$ decreases from $\pi/2$ to $0$. This deviation comes from the non-commuting $\mathcal{L}_h$ and $\mathcal{L}_n$.

We can also define the approximate probabilistic error cancellation map, using the linear expansion method described above in the part of depolarizing noise. The resulting coefficients are
\begin{equation}\label{eq:qAX}
    q_1'=-\frac{\kappa\Delta t}{1-2\kappa\Delta t},\ q_2'=q_3'=0.
\end{equation}
Numerical results are shown in Fig.~\ref{fig:ACAX}. The deviation of the error mitigated results compared with the actual closed dynamics is much larger than the case of Fig.~\ref{fig:ACEX}. This is because of the linear expansion we made, which causes much larger errors than the Trotter errors.

\begin{figure}[t]
\centering
\includegraphics[width=0.48\textwidth]{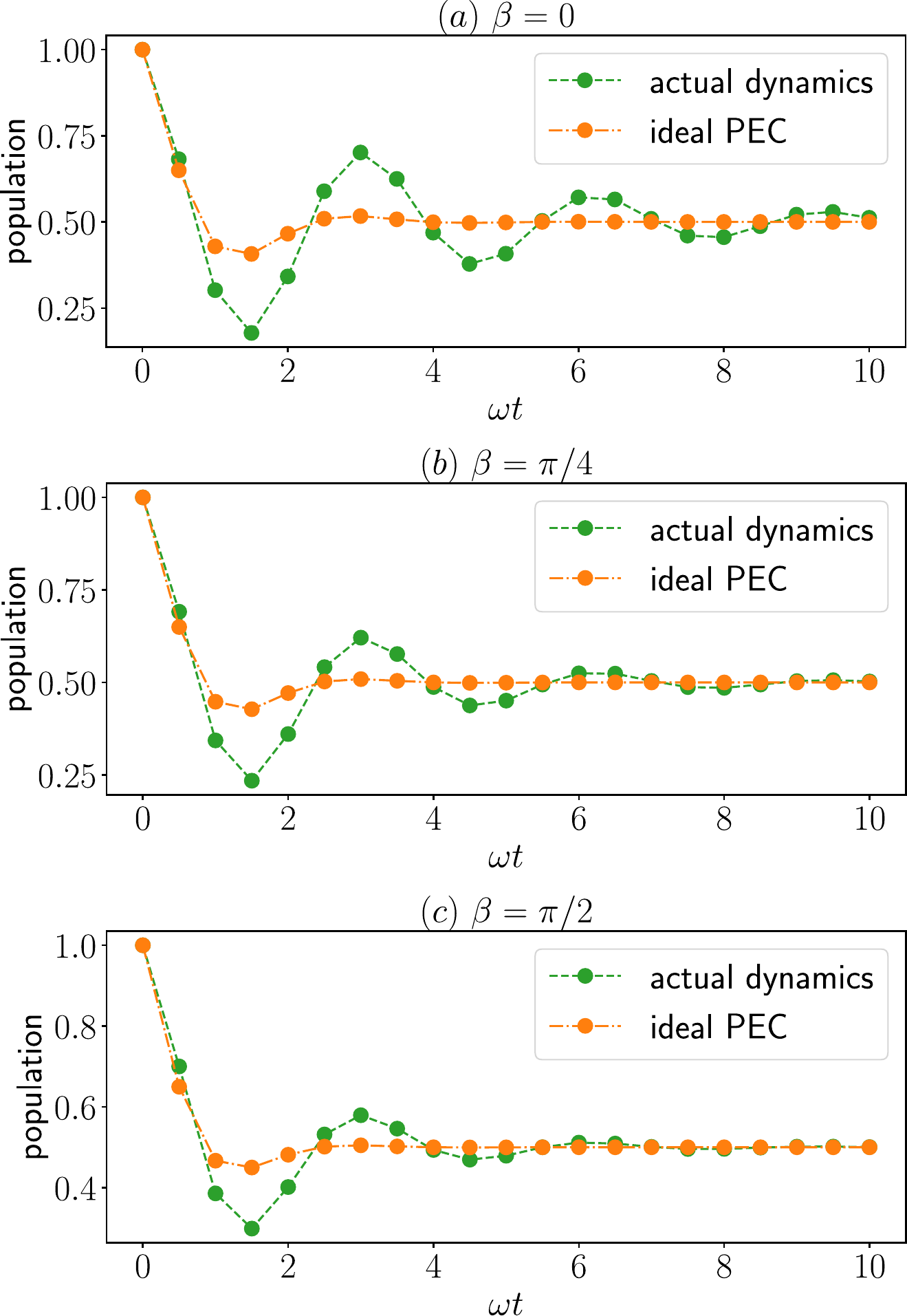}
\caption{Combining step-wise probabilistic error cancellation with a noisy digital quantum computer to simulate open dynamics. We have chosen the parameters as $\omega=1$, $\Delta t=0.5$, $\lambda_1=0.16$, $\lambda_2=0.12$, $\lambda_3=0.2$, $\gamma=0.3$. The green dashed line is the actual open dynamics to be simulated, which is the solution of Eq.~\eqref{eq:GammamasterX}. The orange dot-dashed line is the analytical result (i.e., in the limit of an infinitely large sample size, we have also called it ``ideal implementation") for the approximate error mitigation maps. The parameters $q_0$, $q_1$, $q_2$, $q_3$ in Eq.~\eqref{eq:MsingleQ} follow Eq.~\eqref{eq:doaq}. The difference between the green dashed line and the orange dot-dashed line implies that due to the approximation to the first order of $\gamma\Delta t$, $\lambda_1$, $\lambda_2$ and $\lambda_3$, even in the limit of an infinitely large number of samples, the noise cannot be fully cancelled. (a) $\beta=0$. (b) $\beta=\pi/4$. (c) $\beta=\pi/2$.}\label{fig:DOA}
\end{figure}

\subsection{open dynamics}

Now we will demonstrate the quantum simulation of open dynamics, where the Trotter errors nontrivially depend on the superoperators of the noise part of the open dynamics to be simulated ($\mathcal{L}_d$), the unitary part ($\mathcal{L}_h$), and the device noise ($\mathcal{L}_n$), as argued in Sec.~\ref{sec:general}. Specifically, we consider simulating the following master equation,
\begin{equation}\label{eq:GammamasterX}
    \frac{d\rho(t)}{dt}=-i[\hat{H},\rho(t)]+\gamma \Big(\hat{X}\rho(t)\hat{X}-\rho(t)\Big),
\end{equation}
describing Rabi oscillations with Pauli-X error continuous in time. Same as before, the Hamiltonian $\hat{H}$ is given by Eq.~\eqref{eq:Hamiltonian}. As $\mathcal{L}_d=\gamma(\hat{X}\otimes\hat{X}-\mathds{1}\otimes\mathds{1})$ only commutes with $\mathcal{L}_h$ for $\beta=\pi/2$, in general the exact solution as expressed by Eq.~\eqref{eq:vec_rho} does not have a simple analytical form. Instead, we plot the exact solutions in the green dashed lines in Fig.~\ref{fig:DOA}, for three values of $\beta$. The time evolution of the exited state population is a damped Rabi oscillation, while $\beta=0$ corresponds to the weakest damping and $\beta=\pi/2$ corresponds to the strongest damping.

\begin{figure}[t]
\centering
\includegraphics[width=0.48\textwidth]{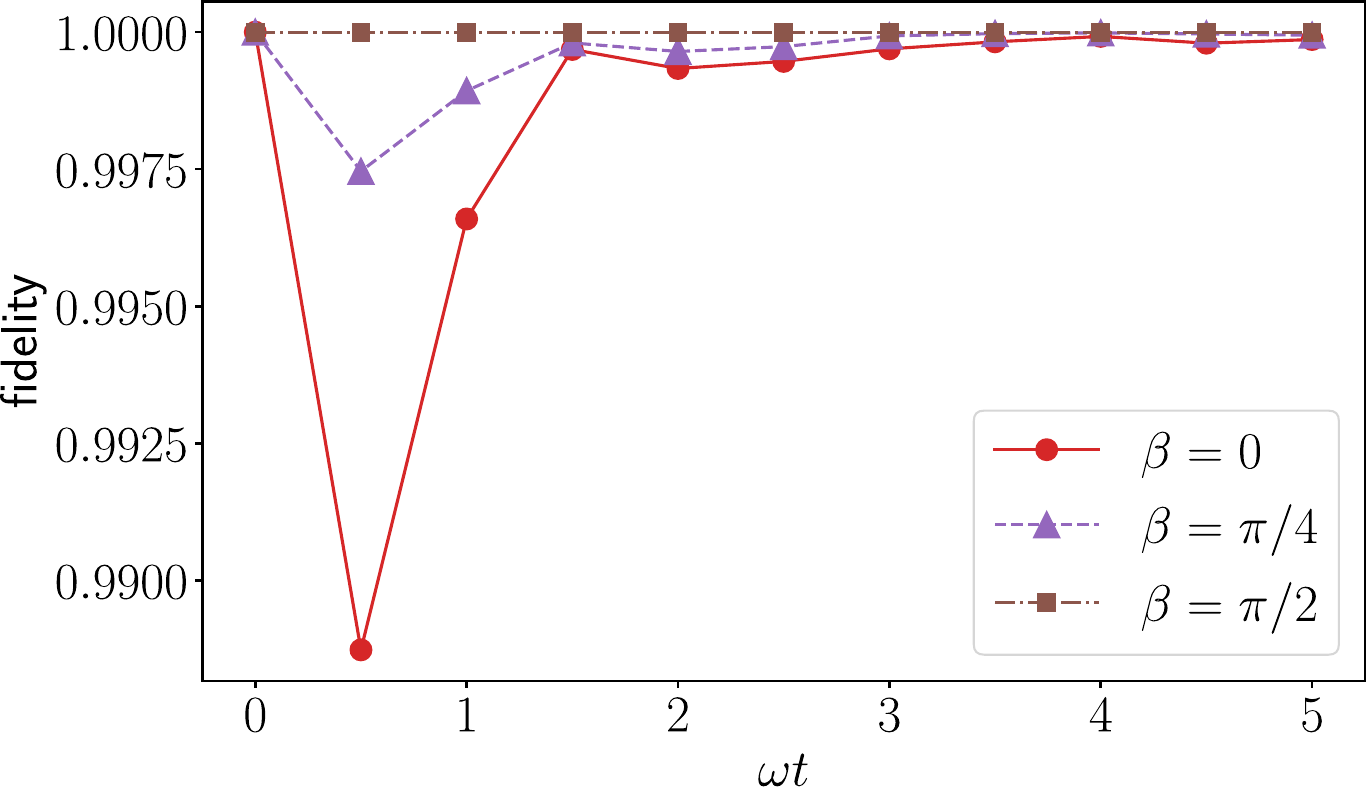}
\caption{Combining step-wise probabilistic error cancellation with a noisy analog quantum computer to simulate open dynamics, assuming a depolarizing noise channel. We have chosen the parameters as $\omega=1$, $\Delta t=0.5$, $\kappa=0.1$, $\gamma=0.3$. The fidelity is defined by comparing the exact solution of Eq.~\eqref{eq:GammamasterX} with the result from the ideal implementation of the approximate error mitigation map with coefficients Eq.~\eqref{eq:qAap}. The red solid line with circle marks is for $\beta=0$. The purple dashed line with triangle marks is for $\beta=\pi/4$. The brown dot-dashed line with square marks is for $\beta=\pi/2$.
}\label{fig:AOA}
\end{figure}

\begin{figure}[t]
\centering
\includegraphics[width=0.48\textwidth]{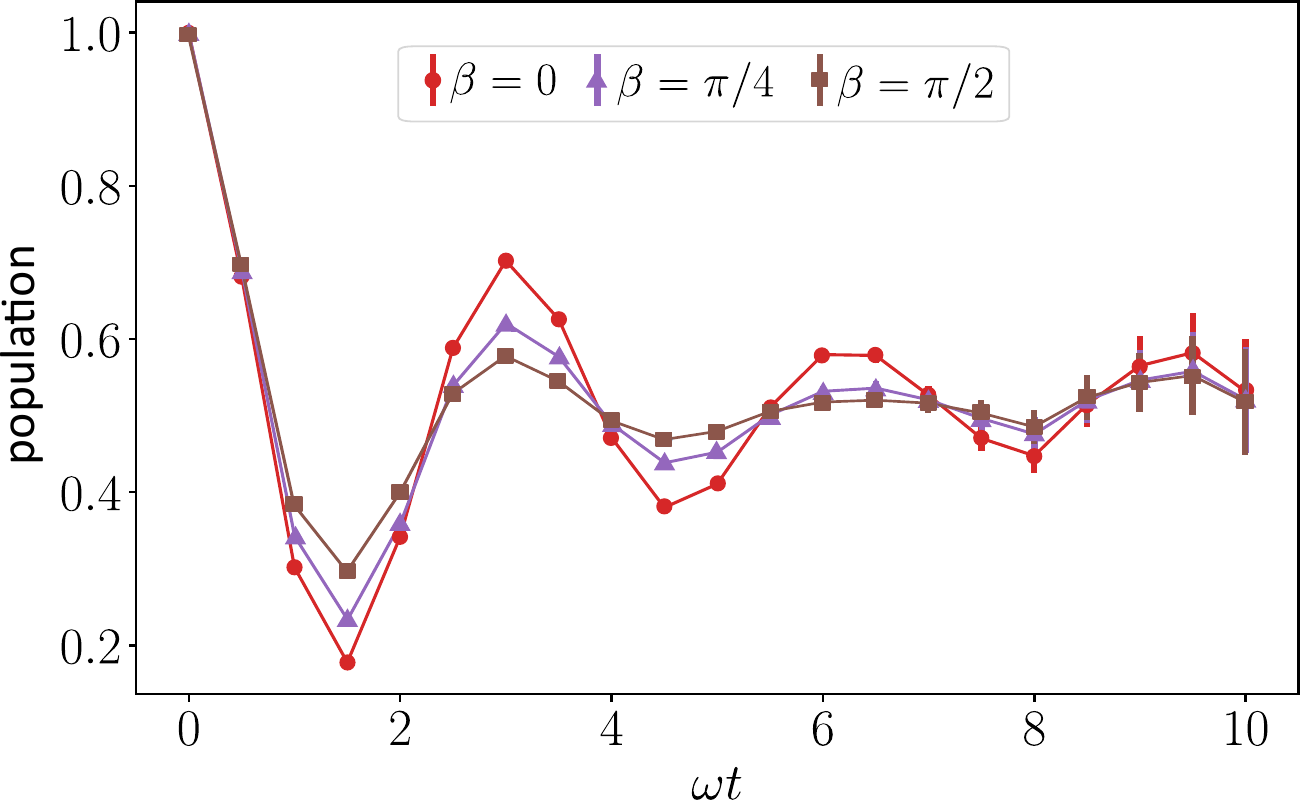}
\caption{Combining step-wise probabilistic error cancellation with a noisy analog quantum computer to simulate open dynamics, assuming a biased Pauli-X noise channel. We have chosen the parameters as $\omega=1$, $\Delta t=0.5$, $\kappa=0.1$, $\gamma=0.3$. The numerical data points are the ensemble averages over $5\times10^6$ samples, and the errorbars correspond to the standard deviations. The parameters $q_0$, $q_1$, $q_2$, $q_3$ in Eq.~\eqref{eq:MsingleQ} correspond to the exact probabilistic error cancellation map, as given in Eq.~\eqref{eq:qACe}. The brown square are data points for $\beta=\pi/2$, the purple triangles are for $\beta=\pi/4$, and the red circles are for $\beta=0$.}\label{fig:AOEX}
\end{figure}

\subsubsection{digital quantum simulation}\label{sec:odq}

We have shown in Sec.~\ref{sec:digitalGeneral} that, if $[\mathcal{L}_d,\mathcal{L}_h]\neq0$, there will be Trotter errors in the quantum simulation. Importantly, this criterion is independent of the device noise superoperator $\mathcal{L}_n$. We choose the device noise channel to follow the same form as in Eq.~\eqref{eq:digitalNoise}. Its vectorized form, $\mathcal{N}=(1-\lambda_1-\lambda_2-\lambda_3)\mathds{1}\otimes\mathds{1}+\lambda_1\hat{X}\otimes\hat{X}+\lambda_2\hat{Y}^{*}\otimes\hat{Y}+\lambda_3\hat{Z}\otimes\hat{Z}$, can be rewritten in the form of $\exp(\mathcal{L}_n\Delta t)$ in order to match Eq.~\eqref{eq:dqsC}, with
\begin{align}\label{eq:ln}
    \mathcal{L}_n&=\kappa_1(\hat{X}\otimes\hat{X}-\mathds{1}\otimes\mathds{1})+\kappa_2(\hat{Y}^*\otimes\hat{Y}-\mathds{1}\otimes\mathds{1})\nonumber\\
    &+\kappa_3(\hat{Z}\otimes\hat{Z}-\mathds{1}\otimes\mathds{1}), 
\end{align}
and coefficients
\begin{align}\label{eq:kappaCoeff}
    &\kappa_1=\frac{1}{4\Delta t}\log\Big(\frac{1-2\lambda_2-2\lambda_3}{(1-2\lambda_1-2\lambda_2)(1-2\lambda_1-2\lambda_3)}\Big),\nonumber\\
    &\kappa_2=\frac{1}{4\Delta t}\log\Big(\frac{1-2\lambda_1-2\lambda_3}{(1-2\lambda_1-2\lambda_2)(1-2\lambda_2-2\lambda_3)}\Big),\nonumber\\
    &\kappa_3=\frac{1}{4\Delta t}\log\Big(\frac{1-2\lambda_1-2\lambda_2}{(1-2\lambda_1-2\lambda_3)(1-2\lambda_2-2\lambda_3)}\Big).
\end{align}
Here we have assumed that the device noise channel is weak, e.g., $\lambda_1+\lambda_2+\lambda_3\leq1/2$. The exact error mitigation superoperator, as defined in Eq.~\eqref{eq:Mdn}, is in the form of Eq.~\eqref{eq:MsingleQ} with coefficients
\begin{align}\label{eq:qOED1}
    q_k=\frac{1}{4}\Big(1-e^{-2\gamma\Delta t}\big(&(-1)^{\delta_{1,k}}\frac{e^{2\gamma\Delta t}}{1-2\lambda_2-2\lambda_3}\nonumber\\
    +&(-1)^{\delta_{2,k}}\frac{1}{1-2\lambda_1-2\lambda_3}\nonumber\\
    +&(-1)^{\delta_{3,k}}\frac{1}{1-2\lambda_1-2\lambda_2}\big)\Big),
\end{align}
where $k=1,2,3$ and $\delta_{i,j}$ is the Kronecker delta.

In Fig.~\ref{fig:DOE}, we show numerical examples of how the commutator $[\mathcal{L}_d,\mathcal{L}_h]$ affects the Trotter errors in the quantum simulation. As we aim at demonstrating the Trotter errors, which are rather small (as clearly seen in the case of Fig.~\ref{fig:ACEX}), we only show the ideal implementation of the error mitigation map, i.e., considering an infinite number of samples. We quantify how close the simulation result is to the exact open dynamics via fidelity, which for qubits is expressed as~\cite{jozsa1994fidelity}
\begin{equation}
    F(\rho_1,\rho_2)=\mathrm{Tr}(\rho_1\rho_2)+2\sqrt{\mathrm{det}(\rho_1)\mathrm{det}(\rho_2)},
\end{equation}
where $\mathrm{det(\cdot)}$ refers to the determinant. As seen in Fig.~\ref{fig:DOE}, only the case of $\beta=\pi/2$ corresponds to simulation results without any error, as for this $\beta$, the Hamiltonian only contains the Pauli $\hat{X}$ operator.

We can also define the approximate error mitigation superoperator by expanding Eq.~\eqref{eq:kappaCoeff} to the first order in $\lambda_1,\lambda_2,\lambda_3$,
\begin{equation}\label{eq:lambdaKappa}
    \kappa_1\approx\frac{\lambda_1}{\Delta t},\ \kappa_2\approx\frac{\lambda_2}{\Delta t},\ \kappa_3\approx\frac{\lambda_3}{\Delta t}.
\end{equation}
The approximate error mitigation superoperator thus has coefficients
\begin{equation}\label{eq:doaq}
    q_1'=\gamma\Delta t-\lambda_1,\ q_2'=-\lambda_2,\ q_3'=-\lambda_3.
\end{equation}
Note that this is the approach taken by Ref.~\cite{guimaraes2023noise}.

In Fig.~\ref{fig:DOA}, we show numerical examples of how the approximation described above affect the simulation results. Although getting the approximated expression of the error mitigation map $\mathcal{M}$ following Eq.~\eqref{eq:doaq} does not require inverting any matrix, the resulting simulation error turns out to be significantly larger than the Trotter error. Most oscillatory features of the damped Rabi oscillation cannot be resolved.

\begin{figure}[t]
\centering
\includegraphics[width=0.48\textwidth]{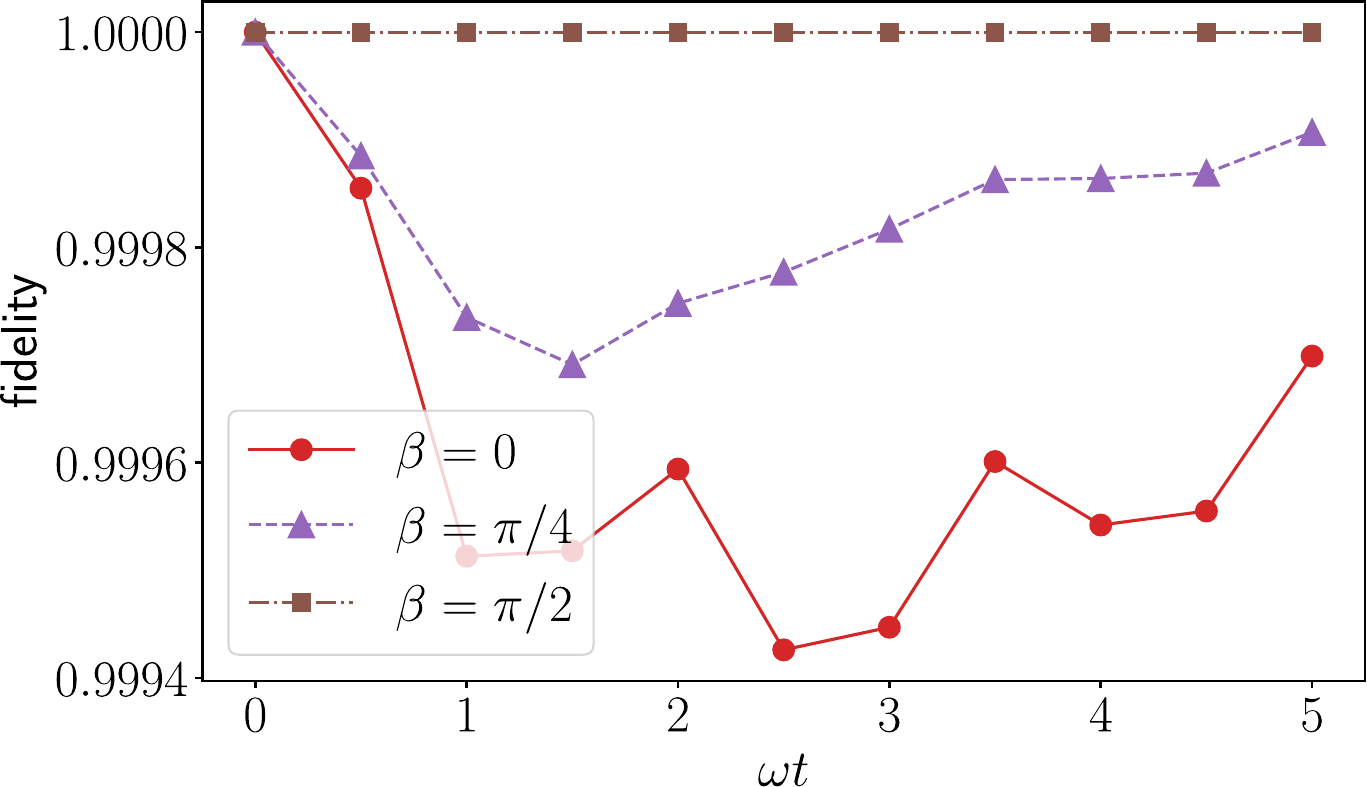}
\caption{Combining step-wise probabilistic error cancellation with a noisy analog quantum computer to simulate open dynamics, assuming a biased Pauli-X noise channel. We have chosen the parameters as $\omega=1$, $\Delta t=0.5$, $\kappa=0.1$, $\gamma=0.3$. The fidelity is defined by comparing the exact solution of Eq.~\eqref{eq:GammamasterX} with the result from the ideal implementation of the approximate error mitigation map with coefficients Eq.~\eqref{eq:qApX}. The red solid line with circle marks is for $\beta=0$. The purple dashed line with triangle marks is for $\beta=\pi/4$. The brown dot-dashed line with square marks is for $\beta=\pi/2$.
}\label{fig:AOAX}
\end{figure}

\subsubsection{analog quantum simulation}\label{sec:analogSub}

Next we focus on noisy analog quantum simulations of the open dynamics. As derived in Sec.~\ref{sec:analog}, the Trotter error now depends on the commutator $[\mathcal{L}_h,\mathcal{L}_d-\mathcal{L}_n]$. We will look into two noise models. The first one is a depolarizing noise model, which will induce Trotter errors. The second one is a biased Pauli-X noise model, which has more Pauli-X noise than a depolarizing noise model. As the open dynamics we aim at simulating only has Pauli-X noise, this noise model will not have Trotter error.

\noindent\textbf{Depolarizing noise.} For the depolarizing noise channel, the system evolves following the master equation Eq.~\eqref{eq:master}. Step-wise probabilistic error cancellation implements the map Eq.~\eqref{eq:Mdn}. In the form of Eq.~\eqref{eq:MsingleQ}, the coefficients are given by
\begin{equation}\label{eq:qADex}
    q_1=\frac{1}{4}(1+e^{4\kappa\Delta t}-2e^{-2\gamma\Delta t+4\kappa\Delta t}),\ q_2=q_3=\frac{1}{4}(1-e^{4\kappa\Delta t}).
\end{equation}

Interestingly, the simulation results in this case are exactly the same as those shown in Fig.~\ref{fig:DOE}. To prove this, we focus on the superoperator applied to the vectorized state at each time step. For the case in Fig.~\ref{fig:DOE}, the superoperator is
\begin{equation}
    e^{(\mathcal{L}_d-\mathcal{L}_n)\Delta t}e^{\mathcal{L}_n\Delta t}e^{\mathcal{L}_h\Delta t}=e^{\mathcal{L}_d\Delta t}e^{\mathcal{L}_h\Delta t}.
\end{equation}
Here the effective $\mathcal{L}_n$ is defined via Eq.~\eqref{eq:ln} and Eq.~\eqref{eq:kappaCoeff}, and we have used the fact that both $\mathcal{L}_d$ and $\mathcal{L}_n$ only contain Pauli terms, therefore they commute with each other. For the analog simulation case considered here, the superoperator at each time step is
\begin{equation}
    e^{(\mathcal{L}_d-\mathcal{L}_n')\Delta t}e^{\mathcal{L}_n'\Delta t+\mathcal{L}_h\Delta t}=e^{\mathcal{L}_d\Delta t}e^{\mathcal{L}_h\Delta t}.
\end{equation}
Here $\mathcal{L}_n'$ is the depolarizing noise channel as written in Eq.~\eqref{eq:lnDp}, and we have used that the depolarizing noise commutes both with $\mathcal{L}_d$ and with $\mathcal{L}_h$. As the open dynamics to be simulated corresponds to the step-wise superoperator $\exp(\mathcal{L}_d\Delta t+\mathcal{L}_h\Delta t)$, Trotter error exists, as illustrated in Fig.~\ref{fig:DOE}.

We can also define the approximate error mitigation map, by expanding Eq.~\eqref{eq:qADex} to the first order in $\kappa\Delta t$ and $\gamma\Delta t$. This results in the coefficients
\begin{equation}\label{eq:qAap}
    q_1'=(\gamma-\kappa)\Delta t,\ q_2'=q_3'=-\kappa\Delta t.
\end{equation}
The numerical results are shown in Fig.~\ref{fig:AOA}. Although the fidelities are close to $1$, the errors are not just Trotter errors, but also the errors from the linear expansion of $q_1$, $q_2$ and $q_3$. Compared with the case in Fig.~\ref{fig:DOA}, the errors here are much smaller, as there is no step similar to the approximation in Eq.~\eqref{eq:lambdaKappa}. Interestingly, the errors here are even smaller than the case of the exact error mitigation map, Fig.~\ref{fig:DOE}, possibly because the Trotter errors and the linear approximation errors cancel with each other.

\noindent\textbf{Biased Pauli-X noise.} Now we consider a biased Pauli-X noise model, such that $[\mathcal{L}_h,\mathcal{L}_d-\mathcal{L}_n]=0$ is satisfied. We assume that the system evolves following the master equation
\begin{align}\label{eq:masterBiasX}
    \frac{d\rho(t)}{dt}=&-i[\hat{H},\rho(t)]+(\gamma+\kappa) \Big(\hat{X}\rho(t)\hat{X}-\rho(t)\Big)\nonumber\\
    &+\kappa \Big(\hat{Y}\rho(t)\hat{Y}-\rho(t)\Big)+\kappa \Big(\hat{Z}\rho(t)\hat{Z}-\rho(t)\Big).
\end{align}
The exact step-wise probabilistic error cancellation superoperator Eq.~\eqref{eq:Mdn}, written in the form of Eq.~\eqref{eq:MsingleQ}, has coefficients that are the same as Eq.~\eqref{eq:qACe}. Numerical results are shown in Fig.~\ref{fig:AOEX}. As there is no Trotter error for all values of $\beta$, the numerical data match the analytical actual dynamics in Fig.~\ref{fig:DOA}.

For the approximate error mitigation map, we can expand Eq.~\eqref{eq:qACe} to the first order in $\kappa\Delta t$ and get
\begin{equation}\label{eq:qApX}
    q_1'=q_2'=q_3'=-\kappa\Delta t.
\end{equation}
The numerical results are shown in Fig.~\ref{fig:AOAX}. The small deviations of the fidelity from $1$ are results of the linear approximations we have made, rather than the Trotter errors.

\section{conclusion and discussion}

We have analyzed the limitations of combining step-wise probabilistic error cancellation with noisy quantum simulations of continuous dynamics beyond the exponentially large sampling overhead. We have pointed out the Trotter errors originating from the superoperators governing the unitary dynamics $\mathcal{L}_h$, the device noise $\mathcal{L}_n$, and the noise part $\mathcal{L}_d$ of the open dynamics to be simulated. These Trotter errors exist even in the limit of an infinite sample size. For a digital quantum simulation, a non-zero commutator $[\mathcal{L}_d,\mathcal{L}_h]$ leads to the Trotter error. For an analog quantum simulation, a non-zero commutator $[\mathcal{L}_d-\mathcal{L}_n,\mathcal{L}_h]$ leads to the Trotter error. Importantly, even for simulating closed dynamics, $\mathcal{L}_d=0$, this Trotter error can still exist for an analog quantum simulation. We have also pointed out that, the commonly used coefficients in the error mitigation map, which do not involve inverting a large matrix, are actually linear approximations of the exact error mitigation map, therefore may induce additional simulation errors. Although dominated by the numerical errors in real quantum devices for now, these errors that we have investigated here put fundamental limitations on the constructions of the simulation methods. It can be expected that as the qualities of the quantum hardware improve, these errors are going to become limiting factors for quantum simulations.

We note that, however, a systematic study as what we have done here is very difficult to go beyond few-qubit toy models and simple noise channels such as Pauli noise channels. How to reduce the Trotter errors in a scalable way remains an open question. One possible way is to design the simulation setup such that, the Trotter error from the non-commuting Hamiltonians of a multi-qubit system partially cancels the Trotter error between the unitary part and the noise part of the dynamics to be simulated. Another possible way may be to remove the time discretization as a middle step, and instead find methods to map one dissipative dynamics to another in a matter that is continuous in time~\cite{kwon2022reversing,harrington2022engineered,donvil2023quantum}, that can be implemented under noisy quantum operations.

\acknowledgements  

We acknowledge the financial support by the Samsung GRC grant and the UK EPSRC grants, EP/Y004752/1 and EP/W032643/1.

\appendix

\section{digital quantum simulation with error probability proportional to the time-step}\label{sec:digital}

\begin{figure}[t]
\centering
\includegraphics[width=0.45\textwidth]{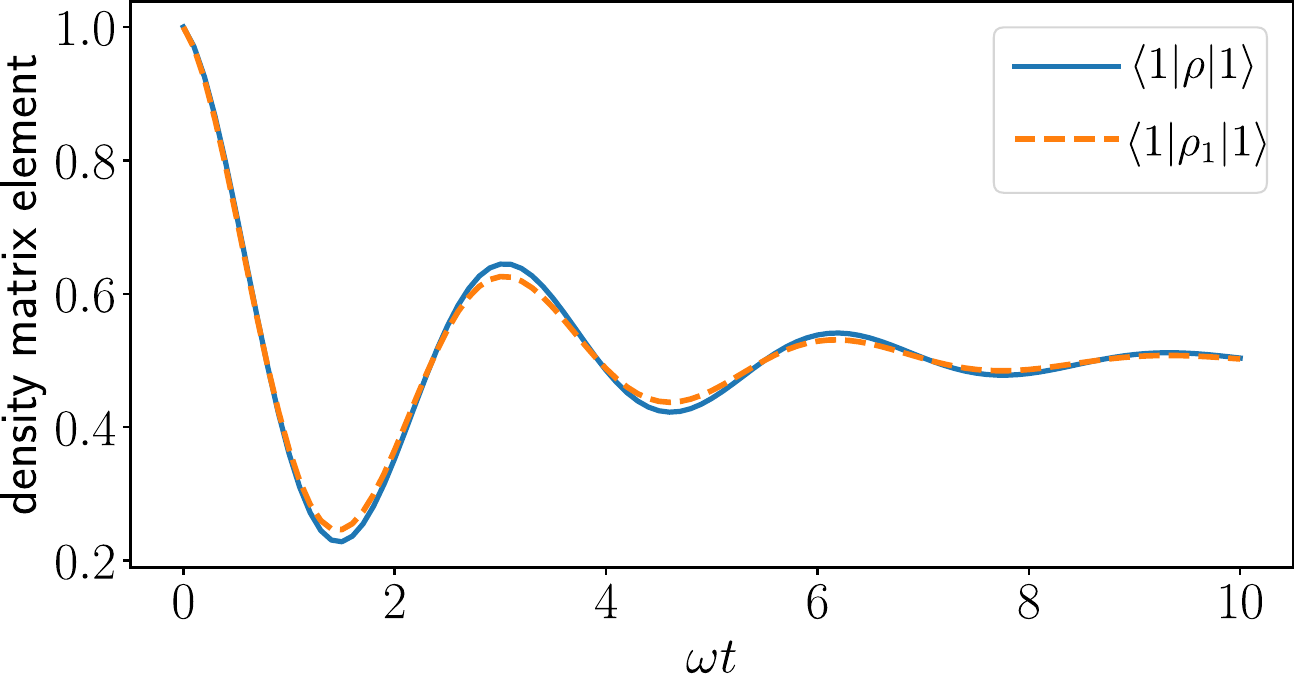}
\caption{Comparison between the exact solution of the master equation [Eq.~\eqref{eq:rho_ee}, blue solid line] and the discretized solution [Eq.~\eqref{eq:rho_ee1}, orange dashed line]. We have chosen the parameters as $\omega=1$, $\kappa=0.1$, $\Delta t=0.5$. Although we have shown $\langle1|\rho_{1}|1\rangle$ as a continuous curve using Eq.~\eqref{eq:kappa}, it is only defined at integer multiples of $\Delta t$.}\label{fig:ana}
\end{figure}

To digitally simulate the unitary dynamics, we take a small time-step $\Delta t$, implement the unitary gate for this time-step on the quantum computer, and apply the unitary gate $N$ times to get the density matrix at time $N\Delta t$. Note that for this toy model, the digitization is somewhat trivial as we can directly write down the closed-form propagator $\hat{U}(t)=\exp(-i\omega t \hat{X})$ which is continuous in time. However, this allows us to focus on the errors in the quantum simulations that are induced solely by dealing with the noise intrinsic to the quantum computer. As a simple example, we suppose that the noise in the quantum computer effectively applies a depolarizing channel after each quantum gate is applied~\cite{guimaraes2023noise}, and the strength of the depolarizing noise channel is proportional to the time-step $\Delta t$. To be specific, in each step the density matrix is transformed in the following way,
\begin{align}\label{eq:dig1}
    \rho_1(t+\Delta t)=&(1-3\kappa\Delta t)e^{-i\omega\Delta t\hat{X}}\rho_1(t)e^{i\omega\Delta t\hat{X}}\nonumber\\
    &+\kappa\Delta t\hat{X}e^{-i\omega\Delta t\hat{X}}\rho_1(t)e^{i\omega\Delta t\hat{X}}\hat{X}\nonumber\\
    &+\kappa\Delta t\hat{Y}e^{-i\omega\Delta t\hat{X}}\rho_1(t)e^{i\omega\Delta t\hat{X}}\hat{Y}\nonumber\\
    &+\kappa\Delta t\hat{Z}e^{-i\omega\Delta t\hat{X}}\rho_1(t)e^{i\omega\Delta t\hat{X}}\hat{Z}.
\end{align}
Here $\rho_1$ denotes the resulting density matrix for this digital simulation, and $\kappa\Delta t\ll 1$ corresponds to the error rate in the depolarizing channel. Note that we have included the factor $\Delta t$ in the definition of the error rate to facilitate a direct comparison in the dissipation rate with Eq.~\eqref{eq:master}. Starting from the initial state $\langle1|\rho(0)|1\rangle=1$, the simulation result is derived as
\begin{equation}\label{eq:rho_ee1}
    \langle1|\rho_{1}(N\Delta t)|1\rangle=\frac{1}{2}\Big(1+(1-4\kappa\Delta t)^N\cos(2\omega N\Delta t)\Big).
\end{equation}
This corresponds to the solution of the master equation \eqref{eq:master} with a time-step-dependent dissipation rate
\begin{align}\label{eq:kappa}
    -\frac{1}{4\Delta t}\log(1-4\kappa\Delta t)\approx\kappa(1+2\kappa\Delta t+\cdots).
\end{align}
In the limit $\Delta t\rightarrow0$, Eq.~\eqref{eq:dig1} is equivalent to Eq.~\eqref{eq:master}, which is also an algebraic way~\cite{ma2022unifying} of deriving Eq.~\eqref{eq:rho_ee}. However, for finite $\Delta t$, Eq.~\eqref{eq:dig1} always corresponds to a stronger dissipation than Eq.~\eqref{eq:master}. As shown in Fig.~\ref{fig:ana}, although the steady state is not changed by this difference, the time evolution before reaching the steady state is changed.

\begin{figure}[t]
\centering
\includegraphics[width=0.45\textwidth]{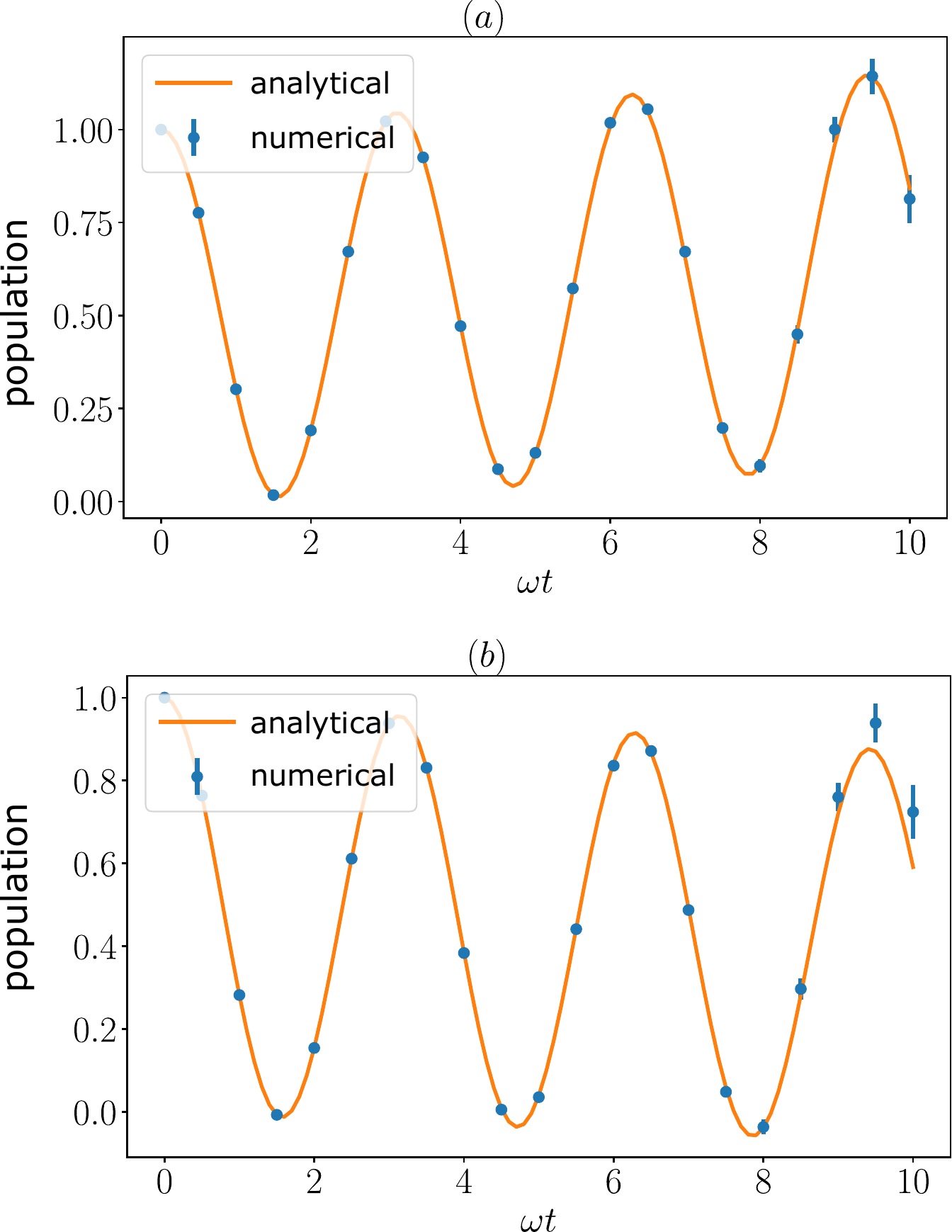}
\caption{Using probabilistic error cancellation to remove the noise, considering imprecise sampling probabilities. We have chosen the parameters as $\omega=1$, $\kappa=0.1$, $\Delta t=0.5$. The blue numerical data points are the ensemble averages over $2\times10^7$ samples, and the errorbars correspond to the standard deviation. The orange lines are the analytical results in the limit of an infinitely large sample size, Eq.~\eqref{eq:rho_ee3}. (a) In each error mitigation step, the probability $\mu'$ of sampling the $\hat{X}$ (or $\hat{Y}$ or $\hat{Z}$) is smaller than the ideal probability $\mu_1$, $\mu'=0.97\mu_1$. This corresponds to $\kappa'\approx0.0041$ and $\xi\approx1.01$, meaning that the trace is increasing with the number of steps. (b)  The probability $\mu'$ of sampling the $\hat{X}$ (or $\hat{Y}$ or $\hat{Z}$) is larger than the ideal probability $\mu_1$, $\mu'=1.03\mu_1$. This corresponds to $\kappa'\approx-0.0042$ and $\xi\approx0.99$, meaning that the trace is decreasing with the number of steps, but instead of a decay, the oscillation amplitude increase in time.}\label{fig:muP1}
\end{figure}

\section{imprecise sampling probabilities}\label{sec:sampling}

In principle, adding probabilistic error cancellation after every time-step in a digital quantum simulation can remove the noise of the quantum computer in a statistical way. However, it has been known that the major disadvantage of quantum error mitigation is that the sampling cost scales exponentially as the number of quantum gates to be applied~\cite{takagi2022fundamental,cai2022quantum}. Here we will point out another issue: When randomly sampling Pauli gates to be applied to the quantum state, deviation in the sampling probabilities may lead to non-physical simulation results. 

The map in Eq.~\eqref{eq:MsingleQ} is non-physical if not all $q_1$, $q_2$ and $q_3$ are non-negative, as shown in the examples in Sec.~\ref{sec:toyModel}. Therefore $\mathcal{M}$ can only be implemented stochastically, together with post-processing, following the description in Sec.~\ref{sec:PEC}. We consider a case where the sampling probabilities are deviated from the ideal values $\mu_1,\mu_2,\mu_3$, but the prefactors are not modified. This can be due to an unknown bias in the sampling process. We take the example of Eq.~\eqref{eq:dig1}, and aim at removing the noise characterized by $\kappa\Delta t$. In order to simplify the expressions, we assume that the probability of applying the $\hat{X}$ (or $\hat{Y}$ or $\hat{Z}$) gate is $\mu'$, which is different from $\mu_1(=\mu_2=\mu_3)$. The implemented map $\mathcal{M}(\rho)$ can be not trace-preserving as we have
\begin{equation}
    \mathrm{Tr}[\mathcal{M}(\rho)]=\frac{(1+2\kappa\Delta t)(1-6\mu')}{1-4\kappa\Delta t}\equiv\xi.
\end{equation}
Applying the error mitigation map $\mathcal{M}(\rho)$ after each step of the noisy digital quantum simulation Eq.~\eqref{eq:dig1} results in the statistically averaged simulation result
\begin{equation}\label{eq:rho_ee3}
    \langle1|\rho(N\Delta t)|1\rangle=\xi^N\frac{1}{2}\Big(1+e^{-4\kappa'N\Delta t}\cos(2\omega N\Delta t)\Big),
\end{equation}
with
\begin{equation}
    \kappa'=\frac{1}{4\Delta t}\log\Big(\frac{1-6\mu'}{(1-4\kappa\Delta t)(1-2\mu')}\Big).
\end{equation}
In addition to the change in the trace described by $\xi$, another non-physical feature is that $\kappa'$ can be negative. The reason of simulation results being non-physical is that, even though the process of randomly sampling a Pauli matrix to apply it to the state is a physical operation that preserves the trace and makes sure that the diagonal elements of the density matrix are non-negative, the post-processing of multiplying the density matrix with  prefactor is not a physical process. Numerical examples are shown in Fig.~\ref{fig:muP1}. Although $\xi$ only slightly deviates from $1$ and if $\kappa'<0$ it is very close to 0, the exponential in $N$ means that the non-physical features grow exponentially in the number of steps, i.e., circuit depth.

\section{explicit expression of the error mitigation superoperator for single qubit Pauli error models}\label{sec:PauliDetails}

In this section, we will give the explicit expression of Eq.~\eqref{eq:MsingleQ}, which is Eq.~\eqref{eq:Mdn} assuming a single qubit under Pauli errors.

The superoperator $\mathcal{L}_d$ describes the noise on the qubit that is the noise part of the open dynamics to be simulated,
\begin{align}\label{eq:ldAll}
    \mathcal{L}_d&=\gamma_1(\hat{X}\otimes\hat{X}-\mathds{1}\otimes\mathds{1})+\gamma_2(\hat{Y}^*\otimes\hat{Y}-\mathds{1}\otimes\mathds{1})\nonumber\\
    &+\gamma_3(\hat{Z}\otimes\hat{Z}-\mathds{1}\otimes\mathds{1}), 
\end{align}
where the error rate $\gamma_k>0$. 
The superoperator $\mathcal{L}_n$ describes the effect of the device noise on the qubit, and its explicit form is given in Eq.~\eqref{eq:ln}. The error mitigation superoperator as defined in Eq.~\eqref{eq:Mdn}, is found to have the form of Eq.~\eqref{eq:MsingleQ} with the following coefficients,
\begin{widetext}
\begin{align}
    q_0&=\frac{1}{4}\Big(1+e^{2(\kappa_1-\gamma_1+\kappa_2-\gamma_2)\Delta t}+e^{2(\kappa_3-\gamma_3+\kappa_1-\gamma_1)\Delta t}+e^{2(\kappa_2-\gamma_2+\kappa_3-\gamma_3)\Delta t}\Big),\\
    q_1&=\frac{1}{4}\Big(1-e^{-2(\gamma_1+\gamma_2+\gamma_3-\kappa_1-\kappa_2-\kappa_3)\Delta t}\big(-e^{2(\gamma_1-\kappa_1)\Delta t}+e^{2(\gamma_2-\kappa_2)\Delta t}+e^{2(\gamma_3-\kappa_3)\Delta t}\big)\Big),\\
    q_2&=\frac{1}{4}\Big(1-e^{-2(\gamma_1+\gamma_2+\gamma_3-\kappa_1-\kappa_2-\kappa_3)\Delta t}\big(e^{2(\gamma_1-\kappa_1)\Delta t}-e^{2(\gamma_2-\kappa_2)\Delta t}+e^{2(\gamma_3-\kappa_3)\Delta t}\big)\Big),\\
    q_3&=\frac{1}{4}\Big(1-e^{-2(\gamma_1+\gamma_2+\gamma_3-\kappa_1-\kappa_2-\kappa_3)\Delta t}\big(e^{2(\gamma_1-\kappa_1)\Delta t}+e^{2(\gamma_2-\kappa_2)\Delta t}-e^{2(\gamma_3-\kappa_3)\Delta t}\big)\Big).
\end{align}
\end{widetext}
It is straightforward to check that $q_0+q_1+q_2+q_3=1$ and $q_0>1/4$. The signs of $q_1$, $q_2$ and $q_3$ depend on the values of the error rates $\gamma_k$ and $\kappa_k$.


\begin{thebibliography}{26}%
\makeatletter
\providecommand \@ifxundefined [1]{%
 \@ifx{#1\undefined}
}%
\providecommand \@ifnum [1]{%
 \ifnum #1\expandafter \@firstoftwo
 \else \expandafter \@secondoftwo
 \fi
}%
\providecommand \@ifx [1]{%
 \ifx #1\expandafter \@firstoftwo
 \else \expandafter \@secondoftwo
 \fi
}%
\providecommand \natexlab [1]{#1}%
\providecommand \enquote  [1]{``#1''}%
\providecommand \bibnamefont  [1]{#1}%
\providecommand \bibfnamefont [1]{#1}%
\providecommand \citenamefont [1]{#1}%
\providecommand \href@noop [0]{\@secondoftwo}%
\providecommand \href [0]{\begingroup \@sanitize@url \@href}%
\providecommand \@href[1]{\@@startlink{#1}\@@href}%
\providecommand \@@href[1]{\endgroup#1\@@endlink}%
\providecommand \@sanitize@url [0]{\catcode `\\12\catcode `\$12\catcode
  `\&12\catcode `\#12\catcode `\^12\catcode `\_12\catcode `\%12\relax}%
\providecommand \@@startlink[1]{}%
\providecommand \@@endlink[0]{}%
\providecommand \url  [0]{\begingroup\@sanitize@url \@url }%
\providecommand \@url [1]{\endgroup\@href {#1}{\urlprefix }}%
\providecommand \urlprefix  [0]{URL }%
\providecommand \Eprint [0]{\href }%
\providecommand \doibase [0]{https://doi.org/}%
\providecommand \selectlanguage [0]{\@gobble}%
\providecommand \bibinfo  [0]{\@secondoftwo}%
\providecommand \bibfield  [0]{\@secondoftwo}%
\providecommand \translation [1]{[#1]}%
\providecommand \BibitemOpen [0]{}%
\providecommand \bibitemStop [0]{}%
\providecommand \bibitemNoStop [0]{.\EOS\space}%
\providecommand \EOS [0]{\spacefactor3000\relax}%
\providecommand \BibitemShut  [1]{\csname bibitem#1\endcsname}%
\let\auto@bib@innerbib\@empty
\bibitem [{\citenamefont {Terhal}(2015)}]{terhal2015quantum}%
  \BibitemOpen
  \bibfield  {author} {\bibinfo {author} {\bibfnamefont {B.~M.}\ \bibnamefont
  {Terhal}},\ }\bibfield  {title} {\bibinfo {title} {Quantum error correction
  for quantum memories},\ }\href@noop {} {\bibfield  {journal} {\bibinfo
  {journal} {Reviews of Modern Physics}\ }\textbf {\bibinfo {volume} {87}},\
  \bibinfo {pages} {307} (\bibinfo {year} {2015})}\BibitemShut {NoStop}%
\bibitem [{\citenamefont {Cai}\ \emph {et~al.}(2022)\citenamefont {Cai},
  \citenamefont {Babbush}, \citenamefont {Benjamin}, \citenamefont {Endo},
  \citenamefont {Huggins}, \citenamefont {Li}, \citenamefont {McClean},\ and\
  \citenamefont {O'Brien}}]{cai2022quantum}%
  \BibitemOpen
  \bibfield  {author} {\bibinfo {author} {\bibfnamefont {Z.}~\bibnamefont
  {Cai}}, \bibinfo {author} {\bibfnamefont {R.}~\bibnamefont {Babbush}},
  \bibinfo {author} {\bibfnamefont {S.~C.}\ \bibnamefont {Benjamin}}, \bibinfo
  {author} {\bibfnamefont {S.}~\bibnamefont {Endo}}, \bibinfo {author}
  {\bibfnamefont {W.~J.}\ \bibnamefont {Huggins}}, \bibinfo {author}
  {\bibfnamefont {Y.}~\bibnamefont {Li}}, \bibinfo {author} {\bibfnamefont
  {J.~R.}\ \bibnamefont {McClean}},\ and\ \bibinfo {author} {\bibfnamefont
  {T.~E.}\ \bibnamefont {O'Brien}},\ }\bibfield  {title} {\bibinfo {title}
  {Quantum error mitigation},\ }\href@noop {} {\bibfield  {journal} {\bibinfo
  {journal} {arXiv preprint arXiv:2210.00921}\ } (\bibinfo {year}
  {2022})}\BibitemShut {NoStop}%
\bibitem [{\citenamefont {Preskill}(2018)}]{preskill2018quantum}%
  \BibitemOpen
  \bibfield  {author} {\bibinfo {author} {\bibfnamefont {J.}~\bibnamefont
  {Preskill}},\ }\bibfield  {title} {\bibinfo {title} {Quantum computing in the
  nisq era and beyond},\ }\href@noop {} {\bibfield  {journal} {\bibinfo
  {journal} {Quantum}\ }\textbf {\bibinfo {volume} {2}},\ \bibinfo {pages} {79}
  (\bibinfo {year} {2018})}\BibitemShut {NoStop}%
\bibitem [{\citenamefont {Daley}\ \emph {et~al.}(2022)\citenamefont {Daley},
  \citenamefont {Bloch}, \citenamefont {Kokail}, \citenamefont {Flannigan},
  \citenamefont {Pearson}, \citenamefont {Troyer},\ and\ \citenamefont
  {Zoller}}]{daley2022practical}%
  \BibitemOpen
  \bibfield  {author} {\bibinfo {author} {\bibfnamefont {A.~J.}\ \bibnamefont
  {Daley}}, \bibinfo {author} {\bibfnamefont {I.}~\bibnamefont {Bloch}},
  \bibinfo {author} {\bibfnamefont {C.}~\bibnamefont {Kokail}}, \bibinfo
  {author} {\bibfnamefont {S.}~\bibnamefont {Flannigan}}, \bibinfo {author}
  {\bibfnamefont {N.}~\bibnamefont {Pearson}}, \bibinfo {author} {\bibfnamefont
  {M.}~\bibnamefont {Troyer}},\ and\ \bibinfo {author} {\bibfnamefont
  {P.}~\bibnamefont {Zoller}},\ }\bibfield  {title} {\bibinfo {title}
  {Practical quantum advantage in quantum simulation},\ }\href@noop {}
  {\bibfield  {journal} {\bibinfo  {journal} {Nature}\ }\textbf {\bibinfo
  {volume} {607}},\ \bibinfo {pages} {667} (\bibinfo {year}
  {2022})}\BibitemShut {NoStop}%
\bibitem [{\citenamefont {Kim}\ \emph {et~al.}(2023{\natexlab{a}})\citenamefont
  {Kim}, \citenamefont {Wood}, \citenamefont {Yoder}, \citenamefont {Merkel},
  \citenamefont {Gambetta}, \citenamefont {Temme},\ and\ \citenamefont
  {Kandala}}]{kim2023scalable}%
  \BibitemOpen
  \bibfield  {author} {\bibinfo {author} {\bibfnamefont {Y.}~\bibnamefont
  {Kim}}, \bibinfo {author} {\bibfnamefont {C.~J.}\ \bibnamefont {Wood}},
  \bibinfo {author} {\bibfnamefont {T.~J.}\ \bibnamefont {Yoder}}, \bibinfo
  {author} {\bibfnamefont {S.~T.}\ \bibnamefont {Merkel}}, \bibinfo {author}
  {\bibfnamefont {J.~M.}\ \bibnamefont {Gambetta}}, \bibinfo {author}
  {\bibfnamefont {K.}~\bibnamefont {Temme}},\ and\ \bibinfo {author}
  {\bibfnamefont {A.}~\bibnamefont {Kandala}},\ }\bibfield  {title} {\bibinfo
  {title} {Scalable error mitigation for noisy quantum circuits produces
  competitive expectation values},\ }\href@noop {} {\bibfield  {journal}
  {\bibinfo  {journal} {Nature Physics}\ ,\ \bibinfo {pages} {1}} (\bibinfo
  {year} {2023}{\natexlab{a}})}\BibitemShut {NoStop}%
\bibitem [{\citenamefont {Van Den~Berg}\ \emph {et~al.}(2023)\citenamefont {Van
  Den~Berg}, \citenamefont {Minev}, \citenamefont {Kandala},\ and\
  \citenamefont {Temme}}]{van2023probabilistic}%
  \BibitemOpen
  \bibfield  {author} {\bibinfo {author} {\bibfnamefont {E.}~\bibnamefont {Van
  Den~Berg}}, \bibinfo {author} {\bibfnamefont {Z.~K.}\ \bibnamefont {Minev}},
  \bibinfo {author} {\bibfnamefont {A.}~\bibnamefont {Kandala}},\ and\ \bibinfo
  {author} {\bibfnamefont {K.}~\bibnamefont {Temme}},\ }\bibfield  {title}
  {\bibinfo {title} {Probabilistic error cancellation with sparse
  Pauli--Lindblad models on noisy quantum processors},\ }\href@noop {}
  {\bibfield  {journal} {\bibinfo  {journal} {Nature Physics}\ ,\ \bibinfo
  {pages} {1}} (\bibinfo {year} {2023})}\BibitemShut {NoStop}%
\bibitem [{\citenamefont {Arute}\ \emph {et~al.}(2020)\citenamefont {Arute},
  \citenamefont {Arya}, \citenamefont {Babbush}, \citenamefont {Bacon},
  \citenamefont {Bardin}, \citenamefont {Barends}, \citenamefont {Bengtsson},
  \citenamefont {Boixo}, \citenamefont {Broughton}, \citenamefont {Buckley}
  \emph {et~al.}}]{arute2020observation}%
  \BibitemOpen
  \bibfield  {author} {\bibinfo {author} {\bibfnamefont {F.}~\bibnamefont
  {Arute}}, \bibinfo {author} {\bibfnamefont {K.}~\bibnamefont {Arya}},
  \bibinfo {author} {\bibfnamefont {R.}~\bibnamefont {Babbush}}, \bibinfo
  {author} {\bibfnamefont {D.}~\bibnamefont {Bacon}}, \bibinfo {author}
  {\bibfnamefont {J.~C.}\ \bibnamefont {Bardin}}, \bibinfo {author}
  {\bibfnamefont {R.}~\bibnamefont {Barends}}, \bibinfo {author} {\bibfnamefont
  {A.}~\bibnamefont {Bengtsson}}, \bibinfo {author} {\bibfnamefont
  {S.}~\bibnamefont {Boixo}}, \bibinfo {author} {\bibfnamefont
  {M.}~\bibnamefont {Broughton}}, \bibinfo {author} {\bibfnamefont {B.~B.}\
  \bibnamefont {Buckley}}, \emph {et~al.},\ }\bibfield  {title} {\bibinfo
  {title} {Observation of separated dynamics of charge and spin in the
  fermi-hubbard model},\ }\href@noop {} {\bibfield  {journal} {\bibinfo
  {journal} {arXiv preprint arXiv:2010.07965}\ } (\bibinfo {year}
  {2020})}\BibitemShut {NoStop}%
\bibitem [{\citenamefont {Rosenberg}\ \emph {et~al.}(2023)\citenamefont
  {Rosenberg}, \citenamefont {Andersen}, \citenamefont {Samajdar},
  \citenamefont {Petukhov}, \citenamefont {Hoke}, \citenamefont {Abanin},
  \citenamefont {Bengtsson}, \citenamefont {Drozdov}, \citenamefont {Erickson},
  \citenamefont {Klimov} \emph {et~al.}}]{rosenberg2023dynamics}%
  \BibitemOpen
  \bibfield  {author} {\bibinfo {author} {\bibfnamefont {E.}~\bibnamefont
  {Rosenberg}}, \bibinfo {author} {\bibfnamefont {T.}~\bibnamefont {Andersen}},
  \bibinfo {author} {\bibfnamefont {R.}~\bibnamefont {Samajdar}}, \bibinfo
  {author} {\bibfnamefont {A.}~\bibnamefont {Petukhov}}, \bibinfo {author}
  {\bibfnamefont {J.}~\bibnamefont {Hoke}}, \bibinfo {author} {\bibfnamefont
  {D.}~\bibnamefont {Abanin}}, \bibinfo {author} {\bibfnamefont
  {A.}~\bibnamefont {Bengtsson}}, \bibinfo {author} {\bibfnamefont
  {I.}~\bibnamefont {Drozdov}}, \bibinfo {author} {\bibfnamefont
  {C.}~\bibnamefont {Erickson}}, \bibinfo {author} {\bibfnamefont
  {P.}~\bibnamefont {Klimov}}, \emph {et~al.},\ }\bibfield  {title} {\bibinfo
  {title} {Dynamics of magnetization at infinite temperature in a heisenberg
  spin chain},\ }\href@noop {} {\bibfield  {journal} {\bibinfo  {journal}
  {arXiv preprint arXiv:2306.09333}\ } (\bibinfo {year} {2023})}\BibitemShut
  {NoStop}%
\bibitem [{\citenamefont {Kim}\ \emph {et~al.}(2023{\natexlab{b}})\citenamefont
  {Kim}, \citenamefont {Eddins}, \citenamefont {Anand}, \citenamefont {Wei},
  \citenamefont {Van Den~Berg}, \citenamefont {Rosenblatt}, \citenamefont
  {Nayfeh}, \citenamefont {Wu}, \citenamefont {Zaletel}, \citenamefont {Temme}
  \emph {et~al.}}]{kim2023evidence}%
  \BibitemOpen
  \bibfield  {author} {\bibinfo {author} {\bibfnamefont {Y.}~\bibnamefont
  {Kim}}, \bibinfo {author} {\bibfnamefont {A.}~\bibnamefont {Eddins}},
  \bibinfo {author} {\bibfnamefont {S.}~\bibnamefont {Anand}}, \bibinfo
  {author} {\bibfnamefont {K.~X.}\ \bibnamefont {Wei}}, \bibinfo {author}
  {\bibfnamefont {E.}~\bibnamefont {Van Den~Berg}}, \bibinfo {author}
  {\bibfnamefont {S.}~\bibnamefont {Rosenblatt}}, \bibinfo {author}
  {\bibfnamefont {H.}~\bibnamefont {Nayfeh}}, \bibinfo {author} {\bibfnamefont
  {Y.}~\bibnamefont {Wu}}, \bibinfo {author} {\bibfnamefont {M.}~\bibnamefont
  {Zaletel}}, \bibinfo {author} {\bibfnamefont {K.}~\bibnamefont {Temme}},
  \emph {et~al.},\ }\bibfield  {title} {\bibinfo {title} {Evidence for the
  utility of quantum computing before fault tolerance},\ }\href@noop {}
  {\bibfield  {journal} {\bibinfo  {journal} {Nature}\ }\textbf {\bibinfo
  {volume} {618}},\ \bibinfo {pages} {500} (\bibinfo {year}
  {2023}{\natexlab{b}})}\BibitemShut {NoStop}%
\bibitem [{\citenamefont {Guimar{\~a}es}\ \emph {et~al.}(2023)\citenamefont
  {Guimar{\~a}es}, \citenamefont {Lim}, \citenamefont {Vasilevskiy},
  \citenamefont {Huelga},\ and\ \citenamefont {Plenio}}]{guimaraes2023noise}%
  \BibitemOpen
  \bibfield  {author} {\bibinfo {author} {\bibfnamefont {J.~D.}\ \bibnamefont
  {Guimar{\~a}es}}, \bibinfo {author} {\bibfnamefont {J.}~\bibnamefont {Lim}},
  \bibinfo {author} {\bibfnamefont {M.~I.}\ \bibnamefont {Vasilevskiy}},
  \bibinfo {author} {\bibfnamefont {S.~F.}\ \bibnamefont {Huelga}},\ and\
  \bibinfo {author} {\bibfnamefont {M.~B.}\ \bibnamefont {Plenio}},\ }\bibfield
   {title} {\bibinfo {title} {Noise-assisted digital quantum simulation of open
  systems},\ }\href@noop {} {\bibfield  {journal} {\bibinfo  {journal} {arXiv
  preprint arXiv:2302.14592}\ } (\bibinfo {year} {2023})}\BibitemShut {NoStop}%
\bibitem [{\citenamefont {Takagi}\ \emph
  {et~al.}(2022{\natexlab{a}})\citenamefont {Takagi}, \citenamefont {Endo},
  \citenamefont {Minagawa},\ and\ \citenamefont {Gu}}]{takagi2022fundamental}%
  \BibitemOpen
  \bibfield  {author} {\bibinfo {author} {\bibfnamefont {R.}~\bibnamefont
  {Takagi}}, \bibinfo {author} {\bibfnamefont {S.}~\bibnamefont {Endo}},
  \bibinfo {author} {\bibfnamefont {S.}~\bibnamefont {Minagawa}},\ and\
  \bibinfo {author} {\bibfnamefont {M.}~\bibnamefont {Gu}},\ }\bibfield
  {title} {\bibinfo {title} {Fundamental limits of quantum error mitigation},\
  }\href@noop {} {\bibfield  {journal} {\bibinfo  {journal} {npj Quantum
  Information}\ }\textbf {\bibinfo {volume} {8}},\ \bibinfo {pages} {114}
  (\bibinfo {year} {2022}{\natexlab{a}})}\BibitemShut {NoStop}%
\bibitem [{\citenamefont {Takagi}\ \emph
  {et~al.}(2022{\natexlab{b}})\citenamefont {Takagi}, \citenamefont {Tajima},\
  and\ \citenamefont {Gu}}]{takagi2022universal}%
  \BibitemOpen
  \bibfield  {author} {\bibinfo {author} {\bibfnamefont {R.}~\bibnamefont
  {Takagi}}, \bibinfo {author} {\bibfnamefont {H.}~\bibnamefont {Tajima}},\
  and\ \bibinfo {author} {\bibfnamefont {M.}~\bibnamefont {Gu}},\ }\bibfield
  {title} {\bibinfo {title} {Universal sampling lower bounds for quantum error
  mitigation},\ }\href@noop {} {\bibfield  {journal} {\bibinfo  {journal}
  {arXiv preprint arXiv:2208.09178}\ } (\bibinfo {year}
  {2022}{\natexlab{b}})}\BibitemShut {NoStop}%
\bibitem [{\citenamefont {Tsubouchi}\ \emph {et~al.}(2022)\citenamefont
  {Tsubouchi}, \citenamefont {Sagawa},\ and\ \citenamefont
  {Yoshioka}}]{tsubouchi2022universal}%
  \BibitemOpen
  \bibfield  {author} {\bibinfo {author} {\bibfnamefont {K.}~\bibnamefont
  {Tsubouchi}}, \bibinfo {author} {\bibfnamefont {T.}~\bibnamefont {Sagawa}},\
  and\ \bibinfo {author} {\bibfnamefont {N.}~\bibnamefont {Yoshioka}},\
  }\bibfield  {title} {\bibinfo {title} {Universal cost bound of quantum error
  mitigation based on quantum estimation theory},\ }\href@noop {} {\bibfield
  {journal} {\bibinfo  {journal} {arXiv preprint arXiv:2208.09385}\ } (\bibinfo
  {year} {2022})}\BibitemShut {NoStop}%
\bibitem [{\citenamefont {Breuer}\ \emph {et~al.}(2002)\citenamefont {Breuer},
  \citenamefont {Petruccione} \emph {et~al.}}]{breuer2002theory}%
  \BibitemOpen
  \bibfield  {author} {\bibinfo {author} {\bibfnamefont {H.-P.}\ \bibnamefont
  {Breuer}}, \bibinfo {author} {\bibfnamefont {F.}~\bibnamefont {Petruccione}},
  \emph {et~al.},\ }\href@noop {} {\emph {\bibinfo {title} {The theory of open
  quantum systems}}}\ (\bibinfo  {publisher} {Oxford University Press on
  Demand},\ \bibinfo {year} {2002})\BibitemShut {NoStop}%
\bibitem [{\citenamefont {Gardiner}\ and\ \citenamefont
  {Zoller}(2004)}]{gardiner2004quantum}%
  \BibitemOpen
  \bibfield  {author} {\bibinfo {author} {\bibfnamefont {C.}~\bibnamefont
  {Gardiner}}\ and\ \bibinfo {author} {\bibfnamefont {P.}~\bibnamefont
  {Zoller}},\ }\href@noop {} {\emph {\bibinfo {title} {Quantum noise: a
  handbook of Markovian and non-Markovian quantum stochastic methods with
  applications to quantum optics}}}\ (\bibinfo  {publisher} {Springer Science
  \& Business Media},\ \bibinfo {year} {2004})\BibitemShut {NoStop}%
\bibitem [{\citenamefont {Milz}\ \emph {et~al.}(2017)\citenamefont {Milz},
  \citenamefont {Pollock},\ and\ \citenamefont {Modi}}]{milz2017introduction}%
  \BibitemOpen
  \bibfield  {author} {\bibinfo {author} {\bibfnamefont {S.}~\bibnamefont
  {Milz}}, \bibinfo {author} {\bibfnamefont {F.~A.}\ \bibnamefont {Pollock}},\
  and\ \bibinfo {author} {\bibfnamefont {K.}~\bibnamefont {Modi}},\ }\bibfield
  {title} {\bibinfo {title} {An introduction to operational quantum dynamics},\
  }\href@noop {} {\bibfield  {journal} {\bibinfo  {journal} {Open Systems \&
  Information Dynamics}\ }\textbf {\bibinfo {volume} {24}},\ \bibinfo {pages}
  {1740016} (\bibinfo {year} {2017})}\BibitemShut {NoStop}%
\bibitem [{\citenamefont {Campaioli}\ \emph {et~al.}(2023)\citenamefont
  {Campaioli}, \citenamefont {Cole},\ and\ \citenamefont
  {Hapuarachchi}}]{campaioli2023tutorial}%
  \BibitemOpen
  \bibfield  {author} {\bibinfo {author} {\bibfnamefont {F.}~\bibnamefont
  {Campaioli}}, \bibinfo {author} {\bibfnamefont {J.~H.}\ \bibnamefont
  {Cole}},\ and\ \bibinfo {author} {\bibfnamefont {H.}~\bibnamefont
  {Hapuarachchi}},\ }\bibfield  {title} {\bibinfo {title} {A tutorial on
  quantum master equations: Tips and tricks for quantum optics, quantum
  computing and beyond},\ }\href@noop {} {\bibfield  {journal} {\bibinfo
  {journal} {arXiv preprint arXiv:2303.16449}\ } (\bibinfo {year}
  {2023})}\BibitemShut {NoStop}%
\bibitem [{\citenamefont {Rossmann}(2006)}]{rossmann2006lie}%
  \BibitemOpen
  \bibfield  {author} {\bibinfo {author} {\bibfnamefont {W.}~\bibnamefont
  {Rossmann}},\ }\href@noop {} {\emph {\bibinfo {title} {Lie groups: an
  introduction through linear groups}}},\ Vol.~\bibinfo {volume} {5}\ (\bibinfo
   {publisher} {Oxford University Press on Demand},\ \bibinfo {year}
  {2006})\BibitemShut {NoStop}%
\bibitem [{\citenamefont {Temme}\ \emph {et~al.}(2017)\citenamefont {Temme},
  \citenamefont {Bravyi},\ and\ \citenamefont {Gambetta}}]{temme2017error}%
  \BibitemOpen
  \bibfield  {author} {\bibinfo {author} {\bibfnamefont {K.}~\bibnamefont
  {Temme}}, \bibinfo {author} {\bibfnamefont {S.}~\bibnamefont {Bravyi}},\ and\
  \bibinfo {author} {\bibfnamefont {J.~M.}\ \bibnamefont {Gambetta}},\
  }\bibfield  {title} {\bibinfo {title} {Error mitigation for short-depth
  quantum circuits},\ }\href@noop {} {\bibfield  {journal} {\bibinfo  {journal}
  {Physical review letters}\ }\textbf {\bibinfo {volume} {119}},\ \bibinfo
  {pages} {180509} (\bibinfo {year} {2017})}\BibitemShut {NoStop}%
\bibitem [{\citenamefont {Sun}\ \emph {et~al.}(2021)\citenamefont {Sun},
  \citenamefont {Yuan}, \citenamefont {Tsunoda}, \citenamefont {Vedral},
  \citenamefont {Benjamin},\ and\ \citenamefont {Endo}}]{sun2021mitigating}%
  \BibitemOpen
  \bibfield  {author} {\bibinfo {author} {\bibfnamefont {J.}~\bibnamefont
  {Sun}}, \bibinfo {author} {\bibfnamefont {X.}~\bibnamefont {Yuan}}, \bibinfo
  {author} {\bibfnamefont {T.}~\bibnamefont {Tsunoda}}, \bibinfo {author}
  {\bibfnamefont {V.}~\bibnamefont {Vedral}}, \bibinfo {author} {\bibfnamefont
  {S.~C.}\ \bibnamefont {Benjamin}},\ and\ \bibinfo {author} {\bibfnamefont
  {S.}~\bibnamefont {Endo}},\ }\bibfield  {title} {\bibinfo {title} {Mitigating
  realistic noise in practical noisy intermediate-scale quantum devices},\
  }\href@noop {} {\bibfield  {journal} {\bibinfo  {journal} {Physical Review
  Applied}\ }\textbf {\bibinfo {volume} {15}},\ \bibinfo {pages} {034026}
  (\bibinfo {year} {2021})}\BibitemShut {NoStop}%
\bibitem [{\citenamefont {Endo}\ \emph {et~al.}(2018)\citenamefont {Endo},
  \citenamefont {Benjamin},\ and\ \citenamefont {Li}}]{endo2018practical}%
  \BibitemOpen
  \bibfield  {author} {\bibinfo {author} {\bibfnamefont {S.}~\bibnamefont
  {Endo}}, \bibinfo {author} {\bibfnamefont {S.~C.}\ \bibnamefont {Benjamin}},\
  and\ \bibinfo {author} {\bibfnamefont {Y.}~\bibnamefont {Li}},\ }\bibfield
  {title} {\bibinfo {title} {Practical quantum error mitigation for near-future
  applications},\ }\href@noop {} {\bibfield  {journal} {\bibinfo  {journal}
  {Physical Review X}\ }\textbf {\bibinfo {volume} {8}},\ \bibinfo {pages}
  {031027} (\bibinfo {year} {2018})}\BibitemShut {NoStop}%
\bibitem [{\citenamefont {Jozsa}(1994)}]{jozsa1994fidelity}%
  \BibitemOpen
  \bibfield  {author} {\bibinfo {author} {\bibfnamefont {R.}~\bibnamefont
  {Jozsa}},\ }\bibfield  {title} {\bibinfo {title} {Fidelity for mixed quantum
  states},\ }\href@noop {} {\bibfield  {journal} {\bibinfo  {journal} {Journal
  of modern optics}\ }\textbf {\bibinfo {volume} {41}},\ \bibinfo {pages}
  {2315} (\bibinfo {year} {1994})}\BibitemShut {NoStop}%
\bibitem [{\citenamefont {Kwon}\ \emph {et~al.}(2022)\citenamefont {Kwon},
  \citenamefont {Mukherjee},\ and\ \citenamefont {Kim}}]{kwon2022reversing}%
  \BibitemOpen
  \bibfield  {author} {\bibinfo {author} {\bibfnamefont {H.}~\bibnamefont
  {Kwon}}, \bibinfo {author} {\bibfnamefont {R.}~\bibnamefont {Mukherjee}},\
  and\ \bibinfo {author} {\bibfnamefont {M.}~\bibnamefont {Kim}},\ }\bibfield
  {title} {\bibinfo {title} {Reversing lindblad dynamics via continuous petz
  recovery map},\ }\href@noop {} {\bibfield  {journal} {\bibinfo  {journal}
  {Physical Review Letters}\ }\textbf {\bibinfo {volume} {128}},\ \bibinfo
  {pages} {020403} (\bibinfo {year} {2022})}\BibitemShut {NoStop}%
\bibitem [{\citenamefont {Harrington}\ \emph {et~al.}(2022)\citenamefont
  {Harrington}, \citenamefont {Mueller},\ and\ \citenamefont
  {Murch}}]{harrington2022engineered}%
  \BibitemOpen
  \bibfield  {author} {\bibinfo {author} {\bibfnamefont {P.~M.}\ \bibnamefont
  {Harrington}}, \bibinfo {author} {\bibfnamefont {E.~J.}\ \bibnamefont
  {Mueller}},\ and\ \bibinfo {author} {\bibfnamefont {K.~W.}\ \bibnamefont
  {Murch}},\ }\bibfield  {title} {\bibinfo {title} {Engineered dissipation for
  quantum information science},\ }\href@noop {} {\bibfield  {journal} {\bibinfo
   {journal} {Nature Reviews Physics}\ }\textbf {\bibinfo {volume} {4}},\
  \bibinfo {pages} {660} (\bibinfo {year} {2022})}\BibitemShut {NoStop}%
\bibitem [{\citenamefont {Donvil}\ \emph {et~al.}(2023)\citenamefont {Donvil},
  \citenamefont {Lechler}, \citenamefont {Ankerhold}, \citenamefont
  {Muratore-Ginanneschi} \emph {et~al.}}]{donvil2023quantum}%
  \BibitemOpen
  \bibfield  {author} {\bibinfo {author} {\bibfnamefont {B.}~\bibnamefont
  {Donvil}}, \bibinfo {author} {\bibfnamefont {R.}~\bibnamefont {Lechler}},
  \bibinfo {author} {\bibfnamefont {J.}~\bibnamefont {Ankerhold}}, \bibinfo
  {author} {\bibfnamefont {P.}~\bibnamefont {Muratore-Ginanneschi}}, \emph
  {et~al.},\ }\bibfield  {title} {\bibinfo {title} {Quantum trajectory approach
  to error mitigation},\ }\href@noop {} {\bibfield  {journal} {\bibinfo
  {journal} {arXiv preprint arXiv:2305.19874}\ } (\bibinfo {year}
  {2023})}\BibitemShut {NoStop}%
\bibitem [{\citenamefont {Ma}\ \emph {et~al.}(2022)\citenamefont {Ma},
  \citenamefont {Pace},\ and\ \citenamefont {Kim}}]{ma2022unifying}%
  \BibitemOpen
  \bibfield  {author} {\bibinfo {author} {\bibfnamefont {Y.}~\bibnamefont
  {Ma}}, \bibinfo {author} {\bibfnamefont {M.~C.}\ \bibnamefont {Pace}},\ and\
  \bibinfo {author} {\bibfnamefont {M.}~\bibnamefont {Kim}},\ }\bibfield
  {title} {\bibinfo {title} {Unifying the s{\o}rensen-m{\o}lmer gate and the
  milburn gate with an optomechanical example},\ }\href@noop {} {\bibfield
  {journal} {\bibinfo  {journal} {Physical Review A}\ }\textbf {\bibinfo
  {volume} {106}},\ \bibinfo {pages} {012605} (\bibinfo {year}
  {2022})}\BibitemShut {NoStop}%
\end{thebibliography}
\end{document}